\documentclass[12pt,preprint]{aastex}

\begin{document}

\title{Large Area Mapping at 850 Microns. 
IV.  Analysis of the Clump Distribution in the Orion B South Molecular Cloud}

\author{Doug Johnstone\altaffilmark{1,2}, 
Henry Matthews\altaffilmark{3}, and George F. Mitchell\altaffilmark{4}}
\altaffiltext{1}
{National Research Council Canada, 
Herzberg Institute of Astrophysics,
5071 West Saanich Rd, Victoria, BC, V9E 2E7, Canada; doug.johnstone@nrc-cnrc.gc.ca}
\altaffiltext{2}{Department of Physics \& Astronomy, University of Victoria,
Victoria, BC, V8P 1A1, Canada}
\altaffiltext{3}
{National Research Council Canada, 
Herzberg Institute of Astrophysics,
Dominion Radio Astrophysical Observatory,
Box 248, Penticton, BC, V2A 6J9, Canada; henry.matthews@nrc-cnrc.gc.ca}
\altaffiltext{4}{Department of Astronomy and Physics, Saint Mary's University, 
Halifax, NS, B3H 3C3, Canada; gmitchell@ap.stmarys.ca}

\slugcomment{Version \today}

\begin{abstract}\noindent

We present results from a survey of a 1300 arcmin$^2$ region of the
Orion B South molecular cloud, including NGC 2024, NGC 2023, and the
Horsehead Nebula (B33), obtained using the Submillimetre
Common-User Bolometer Array (SCUBA) on the James Clerk Maxwell
Telescope. Submillimeter continuum observations at 450\,$\mu$m and
850\,$\mu$m are discussed.  Using an automated algorithm, 57 discrete
emission features (``clumps'') are identified in the 850\,$\mu$m map. The
physical conditions within these clumps are investigated under the
assumption that the objects are in quasi-hydrostatic equilibrium.  The
best fit dust temperature for the clumps is found to be $T_d = 18 \pm
4\,$K, with the exception of those associated with
the few known far infrared sources
residing in NGC 2024. The latter internally heated sources are found
to be much warmer. In the region surrounding NGC 2023, the clump dust
temperatures agree with clump gas temperatures determined from molecular 
line excitation measurements of the CO molecule. The bounding pressure on
the clumps lies in the range 
$\log(k^{-1}\,P\ {\rm cm}^{3}\,{\rm K}^{-1}) = 6.1 \pm 0.3$. 
The cumulative mass distribution is steep at the high mass
end, as is the stellar Initial Mass Function. The distribution
flattens significantly at lower masses, with a turn-over around 3 --
10\,$M_\odot$.
\end{abstract}


\section{Introduction}\label{s_intro}

This paper continues an effort to quantify the necessary
pre-conditions for star formation via observations of the continuum
emission from cold dust at submillimeter wavelengths. Such observations allow
the determination of the mass spectrum and physical characteristics of
the population of cold (typically 10--30\,K), dense concentrations of
dust, or ``clumps'', each of which is expected to be a site of active,
or eventual, star formation. It is also well known that stars form in
groups (see e.g. Clarke et al. 2000, Elmegreen et al. 2000) and that
these in turn originate from a hierarchy of dust clumps for which the
mass spectrum is closely similar to the stellar initial mass function
(IMF; see e.g. Motte et al. 1998; Johnstone et al. 2000b, hereafter Paper II;
Johnstone et al. 2001, hereafter Paper III; Motte et al. 2001). The
recently launched Spitzer Space Telescope will observe many star-forming
regions in the infrared and thus provide information on which dust and
gas clumps contain deeply embedded protostars.  Together, the submillimeter 
continuum dust maps and the infrared images will yield statistics
on the number of prestellar, Class 0, and Class I protostars and 
significantly enhance our understanding of the lifetime in each stage and 
the changes in natal environment with protostellar phase. 

Previous papers in this series on large area mapping at 850\,$\mu$m
have discussed image reconstruction 
techniques (Johnstone et al. 2000a, hereafter Paper I), and 
results for 
the Ophiuchus cloud (Paper II) and the northern part of the 
Orion B region (Paper III).  In the present paper we turn our attention 
to the southern part of the
Orion B molecular cloud (Lynds 1630). This region includes a number of
objects well known to both amateur and professional astronomers. In
particular, the Horsehead Nebula, B33 (see, e.g., Pound et al. 2003),
is seen in silhouette against the bright background of the HII region
IC\,434 near the three bright stars of Orion's Belt. At a distance of
{approximately 400\,pc [Anthony-Twarog (1982) measured a distance
of 390\,pc]} the Orion B South region also contains the
young star clusters NGC\,2023 and NGC\,2024, and as part of the
greater Orion Molecular Cloud is well-known as one of the nearest
extensive regions of active star formation. An optical
image showing most of the region discussed in this paper, and its
relationship with the ionizing stellar system $\sigma$~Orionis, is
given by Abergel et al. (2002).

One of the first systematic searches for dense molecular cores in the
Orion B region was carried out by Lada et al. (1991a; hereafter LBS)
using the CS J = 2--1 transition, for which the effective critical density 
is about $10^4$~cm$^3$ (Evans 1999).  With a beamwidth of 1.8$'$ a total of 
42 dense cores were found to be concentrated in two relatively small regions 
of the Orion B molecular cloud; 19 of these lie within the region surveyed in
the present paper. Seven of the latter objects were subsequently
mapped by Launhardt et al. (1996) at 1300\,$\mu$m wavelength with an
angular resolution of 12$''$, comparable to that of the present
observations with the JCMT. From these surveys it is evident that
active star formation is confined to two rather limited areas within
the Orion B region.

The region has been surveyed at many wavelengths.
{Early 2.2\,$\mu$m observations of a large part of the
molecular cloud made by Lada et al. (1991b) identified four embedded
star clusters, and} subsequent widefield imaging of the stellar
component is available via the 2MASS images (Carpenter 2000). The
distribution of relatively warm dust was mapped by IRAS, although for
our purposes with limited angular resolution, and in far infrared
emission at 138 and 205\,$\mu$m (Mookerjea et al. 2000). The molecular
component has been surveyed by a number of authors, e.g. Miesh and
Bally (1994) in the $^{13}$CO J = 1--0 transition, and by Kramer et
al. (1996) with lower angular resolution in the $^{13}$CO and
$^{12}$CO J = 2--1 and 3--2 transitions.

The Horsehead Nebula in particular has been
extensively investigated. Pound et al. (2003) have obtained an
interferometric image in the CO J = 1--0 transition of the Horsehead
Nebula with an angular resolution of 10$''$. Abergel
et al. (2003) combine mid-infrared ISOCAM data in the LW2 and LW3
bands with CO, $^{13}$CO and C$^{18}$O 1--0 and 2--1 observations to
model the transient heating of the outer edge of the dust cloud by
photodissociating radiation from the relatively nearby O9.5V system
$\sigma$~Orionis. Teyssier et al. (2004) have obtained detailed
observations of simple cyclic and linear hydrocarbons, which they show
to be distributed similarly to CO within the western edge of the
nebula.  Finally, early results from
submillimeter-wavelength mapping of the Horsehead Nebula region have been
reported by Sandell et al. (2001); these latter
archival data have been included in the dataset for the present work,
which covers a total area about 1300 arcmin$^2$ in extent.

\section{Observations and Data Reduction}\label{s_odr}

The data covering Orion B South were obtained between 1997 and 2000
using SCUBA at the JCMT. The JCMT archive, hosted by the
CADC\footnote{The Canadian Astronomy Data Centre (CADC) is operated by
the Dominion Astrophysical Observatory for the National Research
Council of Canada's Herzberg Institute of Astrophysics.}, was used to
find all relevant observations toward the region.  A total of 135
individual scan and jiggle maps were retrieved, which, when combined,
cover a total observed area of 1300 arcmin$^2$. Each individual
3-arcsec cell in the final data array was measured approximately 40
times at 850$\,\mu$m and 90 times at 450$\,\mu$m. Thus, the total time
spent observing each cell was approximately 5 and 11 seconds
respectively. The weather conditions varied significantly during the
many nights of observing. The mean optical depths at
850$\,\mu$m and 450\,$\mu$m were on average quite favorable, 
$\tau_{850} = 0.26 \pm 0.07$ and $\tau_{450} = 1.4 \pm 0.4$.

The raw bolometer data were reduced using the standard SCUBA software
(Holland et al. 1999) to flat-field, extinction correct, and flux
calibrate.  The extinction corrections and flux calibrations as a
function of time were obtained from the JCMT calibrations archive
(Jenness et al. 2002). {The extinction corrections are tabulated
as a function of time based on a comparison of skydip measurements
taken with the JCMT during normal observing and zenith optical depth
measures taken every few minutes at 225\,GHz by the CSO tau meter. The
flux calibration values are tabulated from all calibration
measurements taken during extended periods of time, e.g. semesters,
over which the telescope and instrument were not known to have changed
significantly.  Typical uncertainties in the flux calibration can be
determined from Jenness et al. 2002 (see their Figures 1 and 2).
At 850\,$\mu$m the uncertainty is approximately $\pm 10\,\%$ while at 
450\,$\mu$m the uncertainty is closer to $\pm 25\,\%$.  These values are
consistent with the uncertainties determined by individual observers
using calibrator sources. Most of this uncertainty is due to 
changes in the telescope surface due to temperature and gravity 
during the night and resultant changes in the beam profile, which 
can be quite significant at 450\,$\mu$m.}

\begin{figure}[htp]
\centering
\includegraphics[width=0.75\textwidth,angle=270,clip]{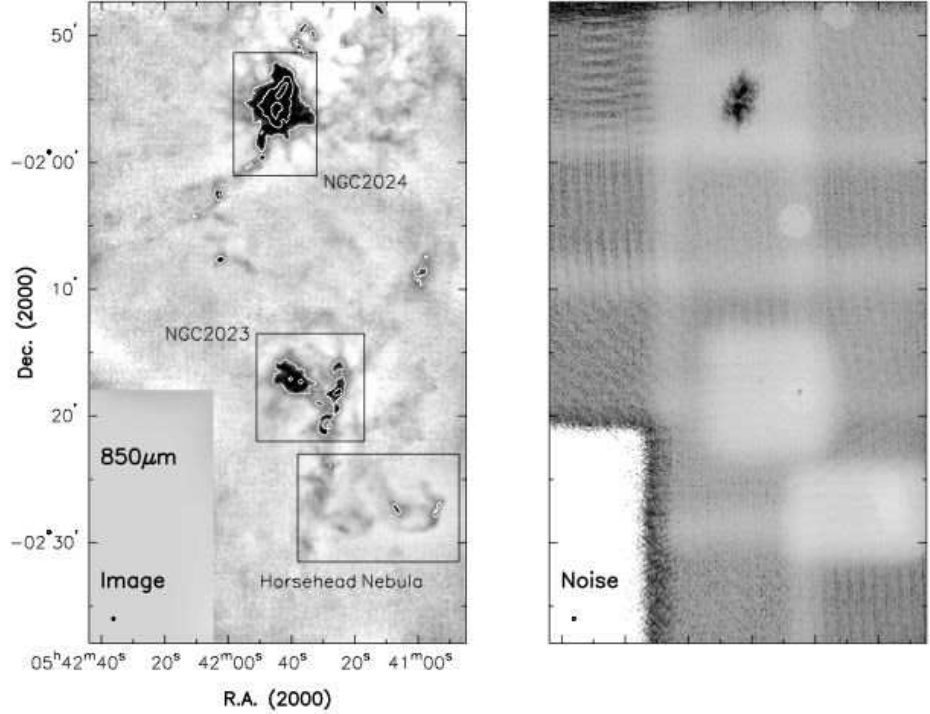}
\caption{The Orion B South region at 850$\,\mu$m. North to south, the
bright star-forming regions are: NGC 2024, NGC 2023, and the
Horsehead.  Left: 850$\,\mu$m emission (greyscale: white to black
represents -0.25 to 0.5 Jy\,bm$^{-1}$, contours are shown for 0.25, 1, 
and 4 Jy\,bm$^{-1}$).  The named boxes indicate the regions displayed in 
Figures \ref{f_obs_env_horse} to \ref{f_obs_env_2024}. The beamsize is 
noted in the lower left of the figure.
Right: 850$\,\mu$m noise map (greyscale: white to
black represents 0 to 0.15 Jy\,bm$^{-1}$).  {In the
noise map, the lower right hand corner has a slightly higher than
typical uncertainty of 0.07 Jy\,pixel$^{-1}$ while the 
Horsehead region directly above has a typical uncertainty of 0.03 
Jy\,pixel$^{-1}$.}  The greater noise in regions
where there is strong signal, such as NGC 2024, is due to
uncertainties in the calibration of individual bolometers.}
\label{f_obs_maps_850}
\end{figure}

\begin{figure}[htp]
\centering
\includegraphics[width=0.75\textwidth,angle=270,clip]{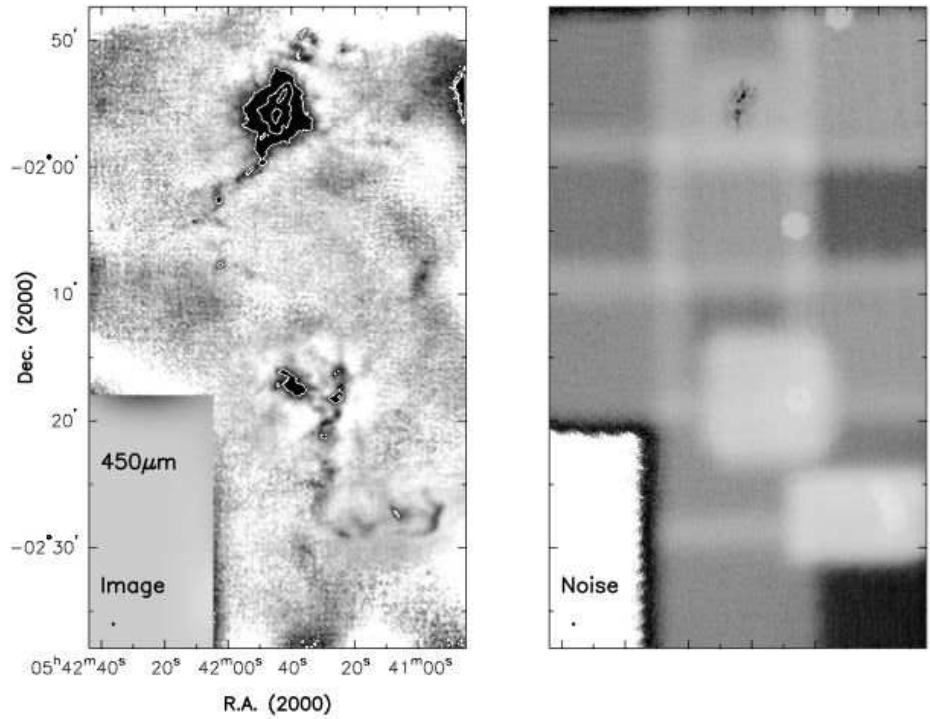}
\caption{The Orion B South region at 450$\,\mu$m, covering the same
area as described in Figure \ref{f_obs_maps_850}.  Left: 450$\,\mu$m
emission (greyscale: white to black represents -1 to 2 Jy\,bm$^{-1}$,
contours at 2, 8, 32 Jy\,bm$^{-1}$). 
The beamsize is noted in the lower left of the figure.
Right: 450$\,\mu$m noise map
(greyscale: white to black represents 0 to 1 Jy\,bm$^{-1}$). {In the
noise map, the lower right hand corner reveals a higher than typical 
uncertainty of 0.9 Jy\,pixel$^{-1}$ while the Horsehead 
region directly above has a typical uncertainty of 0.2
Jy\,pixel$^{-1}$.}}
\label{f_obs_maps_450}
\end{figure}

The data were further reduced and transformed into the final maps
using the matrix inversion technique described in Paper I. This method
has many advantages over the standard facility data reconstruction
procedure, in particular the ability to use all data taken at the
telescope, regardless of chop configuration or mapping procedure, and
the ability to properly weight each individual measurement based on
bolometer and sky noise.  Significant low-amplitude, large-scale
features remain after reconstruction and may be real; however,
artificial features produced by weather variations and bolometer drift
over time will also produce this type of structure since
reconstruction of chopped data amplifies the longer spatial
wavelengths (chop throws range from $20\arcsec$ to $65\arcsec$).  We
filter out these large-scale features by convolving the map with a
large Gaussian profile ($\sigma = 135\arcsec$) and subtracting the
resultant smoothed map from the reconstruction.  In order to minimize
the effect of over-subtraction in regions with bright sources, all
pixels greater than 0.25 Jy\,beam$^{-1}$ {at 850\,$\mu$m and 5
Jy\,beam$^{-1}$ at 450\,$\mu$m} have been removed in the construction
of the smoothed map.  The final, reconstructed 850\,$\mu$m and
450\,$\mu$m maps, after filtering, are shown in Figures
\ref{f_obs_maps_850} and \ref{f_obs_maps_450} along with maps of the
estimated uncertainty at each position. Note that the uncertainty in
the measurement drops significantly in regions where multiple
observations were made, such as the Horsehead region.

\section{Data Analysis: Clump Properties and Distribution}\label{s_clumps}

\subsection{Identification of Clumps}

{We have noted that size scales significantly larger than the
maximum chop throw are unlikely to be correctly registered in the
final images, and such structures have been removed. The data are thus
amenable to an automated clump-finding analysis for
structures of size scales up to approximately $130\arcsec$}.
 
Clumps within the Orion B South region were identified in
substantially the same manner used for Ophiuchus (Paper II) and Orion
B North (Paper III), allowing for a direct comparison of the results.
The flattened 850\,$\mu$m map was searched for structure using a
two-dimensional version of the clump-finding algorithm, {\tt clfind}
(Williams et al. 1994). This algorithm divides the map into individual
clumps by utilizing minimum flux boundaries, assigning each cell in the
map above a limiting threshold to a particular clump. Thus, unlike
Gaussian clump-finding algorithms, the regularity of a clump is not
preset and clumps are allowed to have arbitrary shapes.  A comparison
of this clump-finding algorithm and ``by eye" techniques is discussed
in Paper III, where it is shown that the two techniques produce very
similar results; however, with no observer bias in the present case.

{The typical sensitivity in the raw map is similar to that found
in Paper III, about 40\,mJy per pixel, or 10\,mJy per beam
at 850\,$\mu$m. Comparison can also be made with the
millimeter-wavelength maps of Orion B produced by
Launhardt et al. (1996) which had a sensitivity of approximately 130
mJy per 30\arcsec\ beam. Assuming the dust opacity scales as $\nu^{2}$
the mass sensitivity of the Launhardt observations is about sixteen
times less good than in the present work.
In order to maintain a uniform clump sensitivity across the map, the
typical noise value in the flattened map, approximately 30\,mJy per
pixel, was used as input to {\tt clfind}.}

\begin{figure}[htp]
\centering
\includegraphics[width=1.2\textwidth,clip]{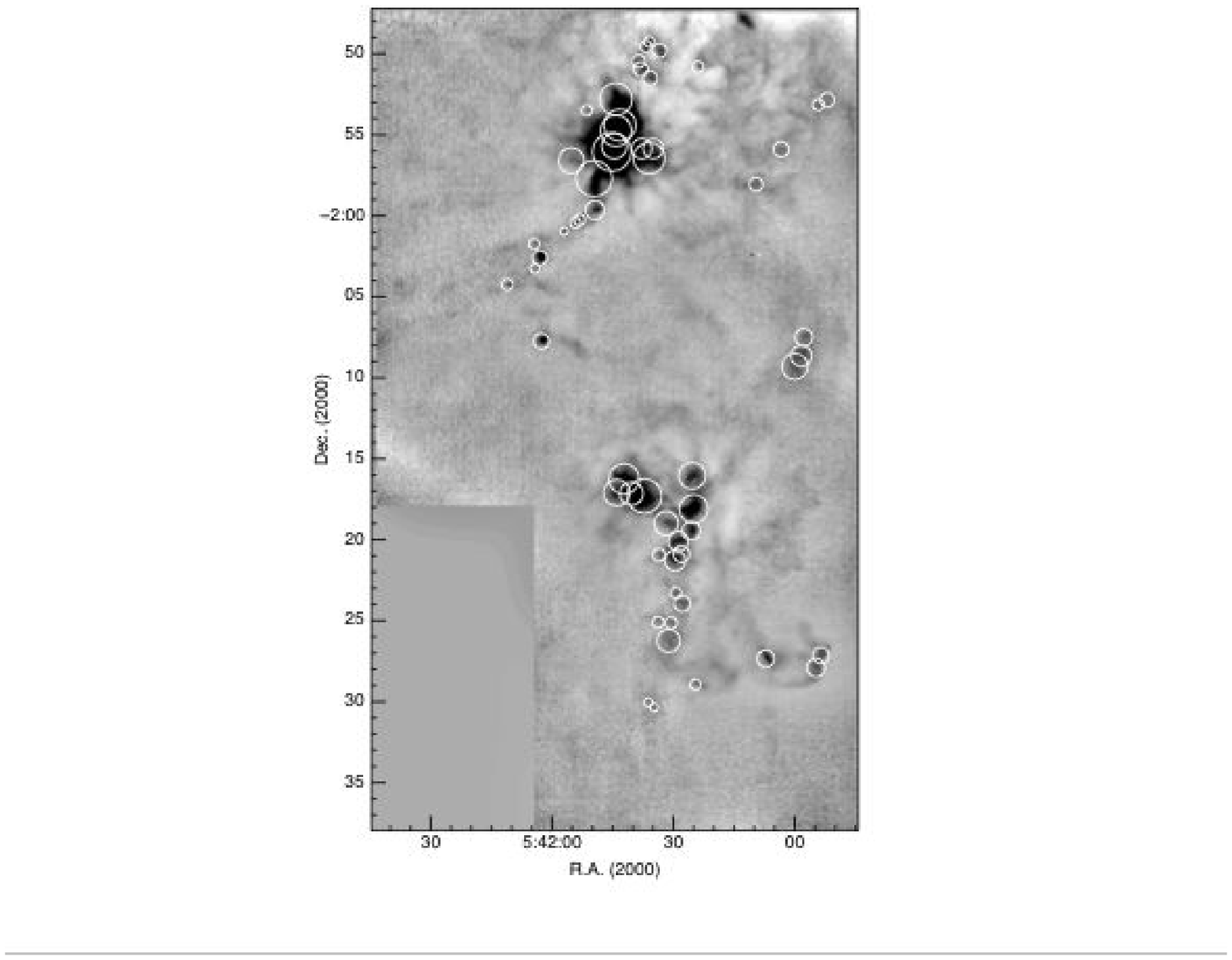}\\
\caption{Location of the submillimeter clumps found in Orion B South 
850$\,\mu$m using an automated procedure (see text). The circle size
approximates the area associated with each clump.}
\label{f_obs_clumps_850}
\end{figure}

Initially, 69 clumps were obtained by the {\tt clfind} routine;
however, twelve were removed due to location within the map, leaving
57 clumps. All removed clumps were near the edges of the map where
residual noise is highest. Following Papers II and III, the size of
each clump was computed by taking the area of each clump and
determining the ``effective radius" of a circle with equivalent
extent.  It should be noted, however, that a number of the clumps
deviate significantly from circular symmetry.  The properties of the
57 clumps are presented in Table \ref{t_850data} and the locations and
relative sizes of the clumps are shown in Figure
\ref{f_obs_clumps_850}. {The total flux for each clump has not
been corrected for the effect of the telescope error beam, in keeping
with previous papers in this series. The error beam at 850\,$\mu$m is
at most $\sim 10\,$\%, much less significant than the calibration
uncertainty.}

An initial estimate of the mass for each clump was computed from the
total flux within the clump boundary under the assumption that the
measured flux was due to thermal emission from optically thin dust
particles {(c.f. equation 1 in Paper III):
\begin{equation} 
M_{\rm clump} = 0.59 \times S_{850}
\left[ \exp\left({17\,{\rm K} \over T_d}\right) -1\right]\,
\left( {\kappa_{850} \over 0.02 {\rm cm}^{2}{\rm g}^{-1}}\right)^{-1} 
\left( {d_{\rm OriB} \over 400\,{\rm pc}}\right)^{2}
\,M_\odot.
\label{e_cm}
\end{equation}
}
Following the results of van
der Tak et al. (1999) we take the opacity per unit mass column density
at 850$\,\mu$m to be $\kappa_{850} = 0.02\,$cm$^2$\,g$^{-1}$, a factor
of two higher than that used in our previous papers. The distance to
Orion B South is assumed to be $d_{850} = 400\,$pc, 
somewhat less than the 450\,pc distance used in Paper III for
Orion B North.  Anticipating the results of the detailed analysis
below, the typical dust temperature for each clump is taken to be $T_d
= 20\,$K, producing clump masses in the range 0.37 to 90$\,M_\odot$.
The four most massive clumps are atypical however; 
they are coincident with known far
infrared sources (Mezger et al. 1988) and have much higher dust and gas
temperatures, $T_d > 50\,K$ (Mangum, Wootten, \& Barsony 1999; Fissel
et al. 2005).  For these four sources the derived masses are
substantially decreased if the higher temperature is adopted, lowering
the maximum clump mass found in Orion B South to less than
30$\,M_\odot$.  The derived mass versus size relation for the clumps
is plotted in Figure \ref{f_obs_mr}, assuming $T_d = 20\,$K for {\it
all} clumps.

In Paper III, the clumps in Orion B North were assumed to have a lower
value of $\kappa$ than used here, and a higher typical dust
temperature, $T_d = 30\,$K.  As well, the distance was taken to be
450\,pc rather than 400\,pc.  To compare derived masses between Paper
III and this paper on a uniform scale, multiply by 0.7 the masses
quoted in Paper III.

\subsection{Clump Distribution}\label{sec:clump-dist}

Within the Orion B South region the three objects B33 (also known as
the Horsehead Nebula), NGC\,2024, and NGC\,2023 contain the large
majority of the star-forming activity and submillimeter-wavelength
emission clumps discussed in this paper.
%
%
We present expanded images of these regions at 850 and 450\,$\mu$m
from the present work, in each case together with a corresponding
optical image from the Digitized Sky Survey (DSS) and the 8-$\mu$m
image obtained by the MSX satellite (Price et al. 2001).

\subsubsection{The Horsehead Nebula}\label{sec:horsehead}

\begin{figure}[htp]
\includegraphics[width=0.63\textwidth,angle=270,clip]{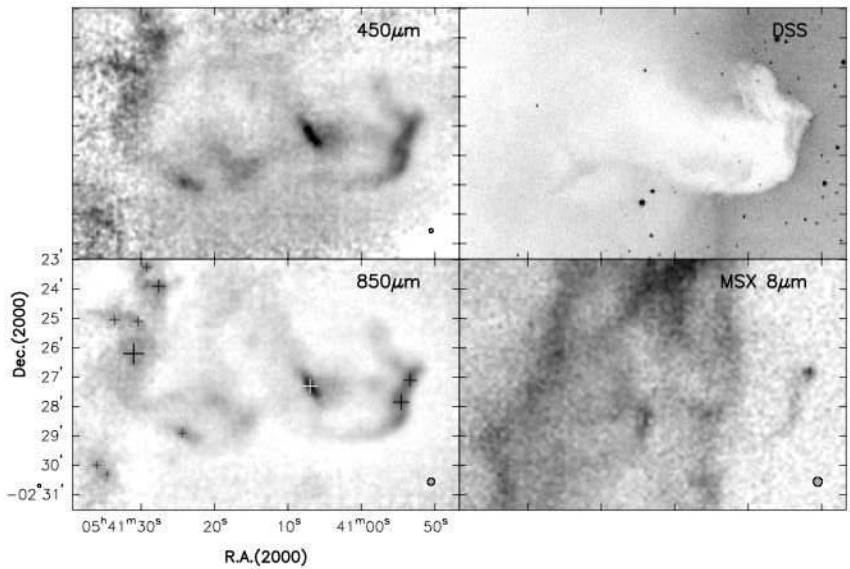}
\caption{\label{f_obs_env_horse} The Horsehead Nebula region (B33). The
panels show the 450 and 850\,$\mu$m emission at upper and lower left
respectively, plotted on intensity scales chosen to illustrate the
structure of the emission to good advantage. Overlaid on the 850\,$\mu$m
map are plus-signs denoting the locations of the submillimeter sources. 
At upper right we show the corresponding 
image from the Digitized Sky Survey (DSS), and at lower right
an image from the 8-$\mu$m data obtained from the MSX database. The
small circle at lower right in each case indicates the respective
beamwidths, except for the DSS data.}
\end{figure}

Of the three regions, only B33 presents aspects
common to all wavelengths (Figure~\ref{f_obs_env_horse}).  The top of
the Horsehead ``neck'' and ``head'', which provide a sharp boundary to
the dark cloud silhouetted against the bright background of the
optical emission nebula IC\,434, is revealed as a region of cold dust
at submillimeter wavelengths. The western edge, heated by $\sigma$
Orionis further to the west (Abergel et al. 2003), appears bright at
8~$\mu$m, and contains a compact feature at mid-IR wavelengths which
could a prestellar object. The latter is not differentiated at sub-mm
wavelengths.

The clump-finding routine identifies two 850\,$\mu$m sources
(054089-02271 and 054091-02278; see Table \ref{t_850data}), both parts
of the ``horse's mane'', in this region, but neither are specifically
associated with the compact mid-IR feature noted above.  Both clumps
have moderately cool submillimeter-wavelength dust temperatures (Table
\ref{t_450850}). A third bright dust clump (054112-02273) $3'$ to the
east, within the ``horse's neck'', has no obvious counterparts at
mid-IR or optical wavelengths, but is amongst the warmer sources based
on the submillimeter data (see Table \ref{t_450850}).  Other
submillimeter sources further to the east are parts of a clumpy ridge
of emission extending south from NGC\,2023. Mid-IR emission possibly
associated with this ridge is displaced slightly to the west, and
likely results from heating by $\sigma$~Orionis.

\subsubsection{NGC\,2023}\label{sec:n2023}
\begin{figure}[htp]
\includegraphics[width=0.85\textwidth,angle=0,clip]{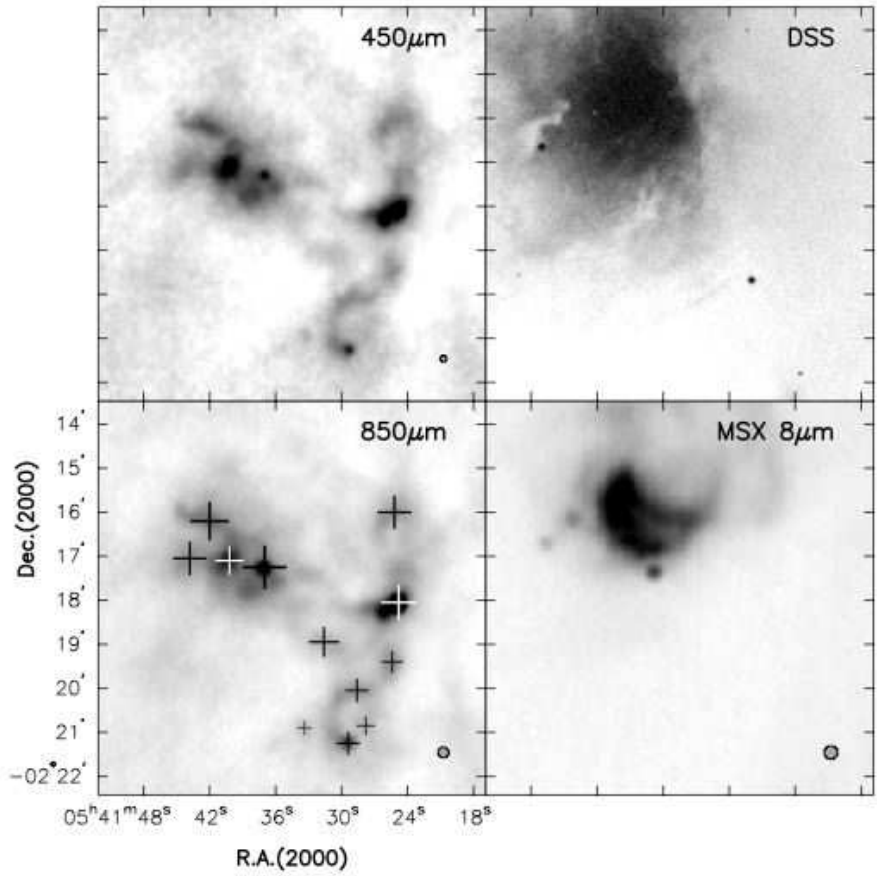}
\caption{\label{f_obs_env_2023} The NGC\,2023 region at submillimeter
(left), optical (upper right) and mid-infrared (lower right)
wavelengths. Other details are as for Figure~\ref{f_obs_env_horse}.}
\end{figure}

Optical and mid-IR images of NGC\,2023 bear no resemblance to the
submillimeter-wavelength images (see Figure~\ref{f_obs_env_2023}), and very
little to each other. The optical image is dominated by a bright
reflection nebula with some patchy absorption by foreground dust. The
illuminating star for the reflection nebula is HD\,37903; heated dust
gives rise to the structure shown the 8-$\mu$m MSX image and also in
ISOCAM data (Abergel et al. 2002). The submillimeter-wavelength data however
reveal a strongly clumped structure containing 12 individual
components. Comparison with the LBS dataset shows that three CS clumps 
lie within NGC\,2023.  LBS\,34 can be identified with the submillimeter clump
054162-02172, while 054142-02160 is likely to be related to LBS\,36.
Interestingly, no LBS clump is identified with the brightest continuum
source in NGC\,2023, 054141-02180; this object is known to harbor a
very cold Class~0 source with a well-collimated molecular outflow (see
Reipurth et al. 2004, and references therein). The remaining LBS clump
within the field in Figure~\ref{f_obs_env_2023}, LBS\,39, is not
clearly associated with compact submillimeter continuum emission.

\subsubsection{NGC\,2024}\label{sec:n2024}
\begin{figure}[htp]
\includegraphics[width=0.85\textwidth,angle=0,clip]{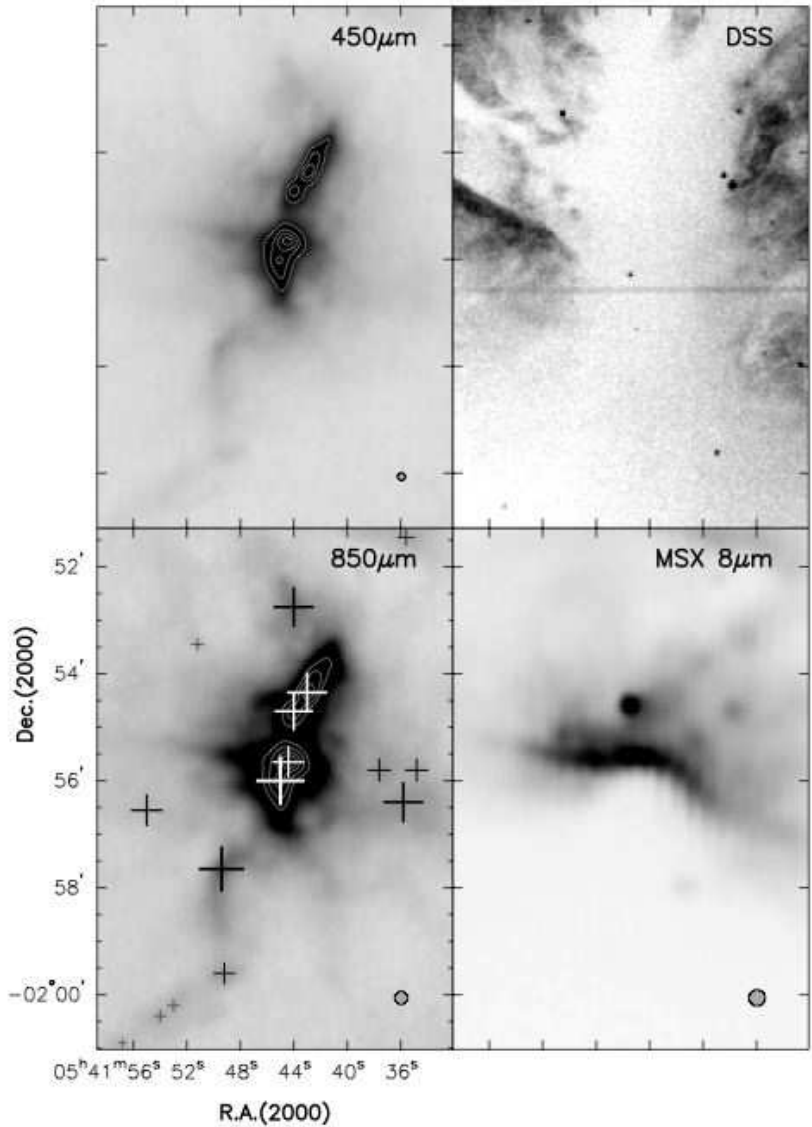}
\caption{\label{f_obs_env_2024}The NGC\,2024 region at submillimeter
(left), optical (upper right) and mid-infrared (lower right)
wavelengths. The horizontal stripe in the DSS image is due to a bright
star outside the frame shown here. Vertical banding in the MSX image
is an artifact. Additional contours for the 450\,$\mu$m image are 30,
45, 60 and 75 Jy/beam area, and for the 850\,$\mu$m image are 4, 6, 9,
12 and 15 Jy/beam area. Other details are as for
Figure~\ref{f_obs_env_horse}.}
\end{figure}

\begin{figure}[htp]
\includegraphics[width=0.65\textwidth,angle=90,clip]{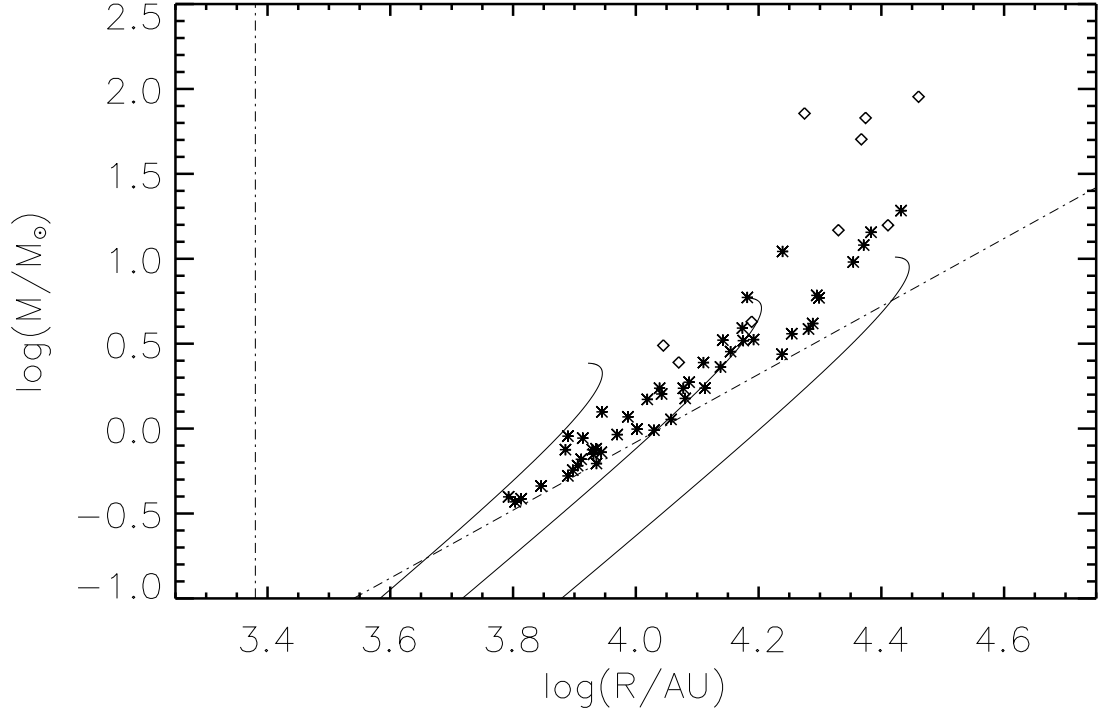}
\caption{\label{f_obs_mr}
Derived mass for each of the 57 clumps, assuming $T_d = 20\,$K, 
versus the effective radius. {Note that the effective radius is derived
from the extent of the clump as determined by {\tt clfind} (see text) and
the largest uncertainty in the mass of the clump is the 
flux calibration, which
is accurate to about twenty percent.} Also plotted (dashed-dotted)
are the minimum size which a clump might have (resolution limit) and the 
minimum mass that a clump must have for a given radius such that the 
clump-finding routine can recognize it (approximately 4 $\sigma$ above the
background).  The symbols denoting each
clump are discussed in Figure \ref{f_obs_con}. The three curves
from left to right denote the mass-radius relation for Bonnor-Ebert
spheres with 20\,K internal temperatures, an additional equal internal pressure
component due to turbulence, and external pressures $k^{-1}\,P = 0.3, 1.0, 3.3
\times 10^6\,$K\,cm$^{-3}$ (see text).}
\end{figure}

As a major star formation region NGC\,2024 has been extensively
studied. As shown in Figure~\ref{f_obs_env_2024} (upper right panel)
the optical appearance is of an extensive ionized region overlaid with
a dense foreground dust lane; the visual extinction toward the
deeply-embedded object IRS\,2 (Grasdalen, 1974) has recently
been determined as $\ge27$ mag at 2.2\,$\mu$m (Lenorzer et al.
2004). Although IRS\,2 contributes significantly to the ionization,
it was known for some time that this star did not meet the O9-O9.5
stellar equivalent required. Recent work by Bik et al. (2003) suggests
that a second fainter star, IRS\,2b, only 5\arcsec\ from IRS\,2,
actually provides the necessary ionization. Centimeter-wave aperture
synthesis continuum observations of this region (Barnes et al. 1989) 
show an HII region with
a sharply-delineated ``bay'' to the south. This structure cannot be
seen at infrared wavelengths, but is replicated in the mid-infrared
emission at 8\,$\mu$m (lower left panel,
Figure~\ref{f_obs_env_2024}).

The submillimeter wavelength observations of NGC\,2024 (see
Figure~\ref{f_obs_env_2024}) show a structure consisting of two major
clumps of dust emission bordered by a number of filamentary
structures. Earlier millimeter-wave observations by Launhardt et al. (1996)
reveal that the two major clumps of emission, as well as the CS clump LBS\,33,
coincide with this structure. At 633~M$_\odot$, this is the most massive
molecular clump found by Lada et al. (1991) in the Orion B South complex.

The SCUBA observations in this paper identify a total of 16 clumps
within the region shown in Figure~\ref{f_obs_env_2024}, including the
four brightest sources in the map.  These objects collectively
and individually represent by far the most massive concentration of
clumped dust within the entire Orion B South region. Notably, all of these
objects have been assigned dust temperatures of 50~K, much higher than
the mean dust temperature in the larger region, on the grounds
that the 850 and 450\,$\mu$m fluxes are inconsistent with a modified
blackbody model. Molecular line observations (Fissel et al. 2005) confirm
that the clumps are warm.

\section{Modeling Clumps as Bonnor-Ebert Spheres}

The nature of the submillimeter-wavelength clumps found in nearby molecular
clouds is at present quite controversial. Whether the clumps are
long-lived, slowly evolving entities, perhaps the result of a gradual
release of magnetic support (Mestel \& Spitzer 1956; Shu 1983; Nakano
1984; Shu et al. 1987; Mouschovias \& Ciolek 1999; Basu \& Ciolek
2004), or transient turbulent structures (Scalo 1985; Mac\,Low \& Klessen
2004 and references therein) has yet to be definitively settled.  The
recently launched Spitzer Space Telescope will play an important role
by determining which clumps have embedded protostars. Regardless,
molecular line observations show that the line widths of such dense cores
are narrow, indicating that non-thermal motions do not overwhelm the
dynamics (Barranco \& Goodman 1998; Goodman et al. 1998), and implying
that equilibrium models should reasonably reproduce the clump
conditions.

In Papers II and III, physical conditions for the clumps were
estimated through modeling the observations as Bonnor-Ebert spheres
(Ebert 1955; Bonnor 1956; Hartmann 1998). {In particular, the
observable size $R\arcsec_{\rm eff}$, the flux $S_{850}$, and 
degree of concentration $C$ for a Bonnor-Ebert sphere map directly 
to the physical mass, temperature, and bounding pressure of the object. 
The concentration is found from the observable size, total flux, peak
flux ${F}_{850}$, and the FWHM beamsize $B\arcsec$ 
(c.f. equation 2 in Paper III):
\begin{equation}
C = 1 - { 1.13\,{B\arcsec}^2\,S_{850} \over 
          \pi\,{R\arcsec_{\rm eff}}^2\,{F}_{850}}. 
\end{equation}
} 

\begin{figure}[htp]
\includegraphics[width=1.0\textwidth,angle=90,clip]{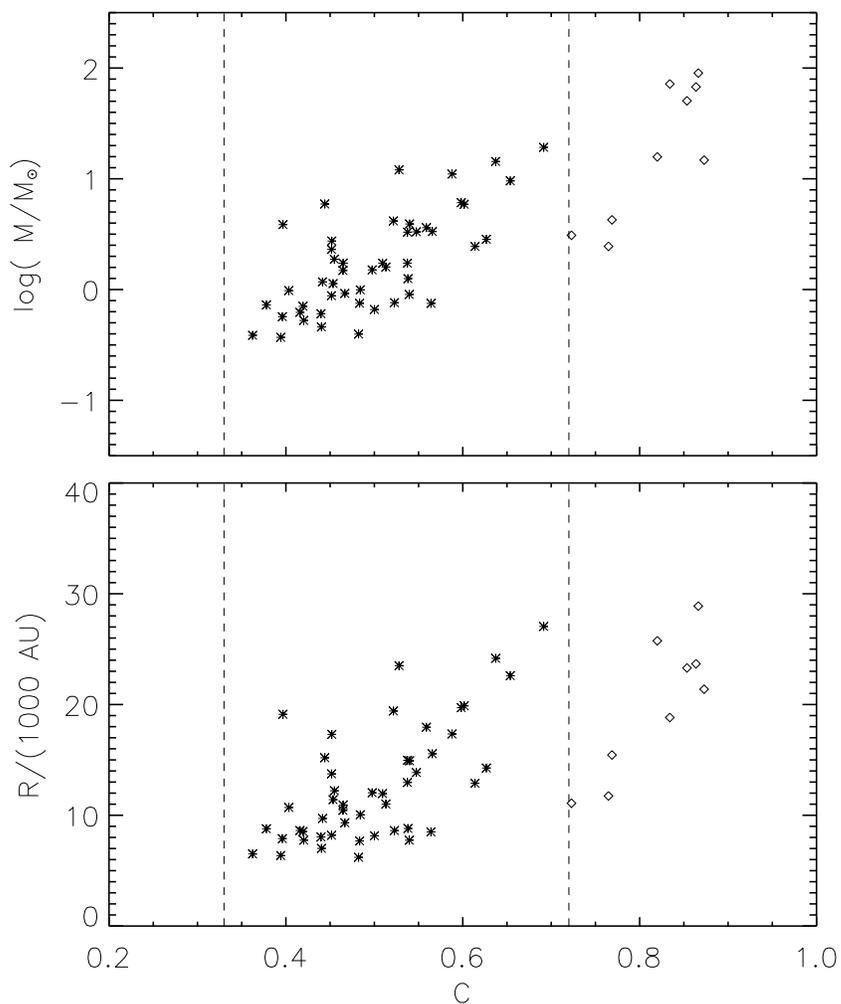}
\caption{\label{f_obs_con}
Derived concentration for each of the 57 clumps in the Orion B South
region (see text). The minimum concentration for a constant density
low-mass Bonnor-Ebert sphere is $C=0.33$, while the maximum
concentration beyond which collapse occurs is $C=0.72$. Clumps with $C
> 0.72$ are denoted by diamonds. (Top) Derived mass of the clump,
assuming $T_d \ 20\,$K,
vs. concentration. (Bottom) Derived radius of the clump
vs. concentration.  }
\end{figure}

A number of important assumptions are required when
applying this technique to available
submillimeter-wavelength data. First, the conversion from flux to
temperature requires knowledge of the dust opacity (see above) and the
assumption that the dust and gas temperatures are related. Given the
high densities for these clumps ($n > 10^4\,$cm$^{-3}$) it is
reasonable to assume the dust and gas temperatures are similar and
thus we take $T_d = T_g$.  In a departure from our earlier models, we
assume also that an additional turbulent component of pressure exists
within the clumps, equal to the thermal pressure at every
location. This change in our analysis is motivated by observations by
Goodman et al.\ (1998), where no clumps were found to have less than
approximately fifty percent turbulent support using dense gas tracer
molecules\footnote{We note, however, that the level of turbulent
support may be strongly dependent on the region. Recent results by
Tafalla (private communication; see also Tafalla et al. 2004) show
that in Taurus the fraction of pressure support due to turbulent
motions is much less than fifty percent.}.  {Given that the bulk
of the molecular cloud has strongly non-thermal support, we take the
transition case, equal contributions from both thermal and
non-thermal, as an approximation for the clump support.}  Finally,
determination of both the clump boundary (i.e. size) and degree of
concentration (i.e. the importance of self-gravity) are complicated by
the 
limited angular resolution and low sensitivity of the
observations. Most of the clumps, however, lie well below the
concentration at which clumps become unstable (see Figure
\ref{f_obs_con}) and thus are well represented by almost constant
density models.  For these cases, determination of the physical
properties, mass, temperature, and bounding pressure are {\it not}
strongly dependent on the determination of the exact value of the
central concentration.  The clumps with the more extreme
concentrations, located in the unstable part of the diagram, are
marked with diamonds in Figure \ref{f_obs_con}.  The known far
infrared sources lie among this group.

\begin{figure}[htp]
\includegraphics[width=0.65\textwidth,angle=90,clip]{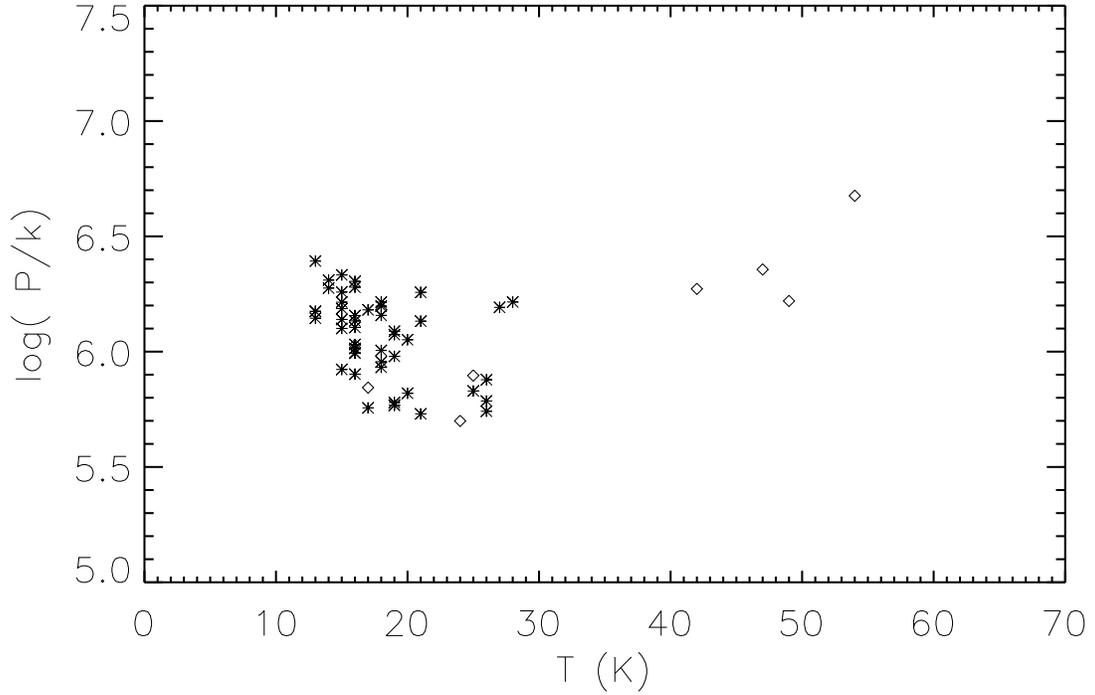}
\caption{\label{f_obs_be} 
Results of determining the physical parameters of clumps using the
assumption that they are well represented by Bonnor-Ebert spheres with
measured concentrations. Plotted are the external, confining pressure
vs.\ the internal temperature (assuming an equal contribution from
turbulent support).  The typical uncertainty in the derived physical
parameters of an individual clump is similar to the spread in the
distribution of points.  The symbols are the same as in Fig.\
\ref{f_obs_con}.  }
\end{figure}

Applying the Bonnor-Ebert sphere analysis used in Papers II and III,
the physical conditions for each clump are plotted in Figure
\ref{f_obs_be}.  Excluding the extreme concentration sources (marked
by diamonds), the typical clump temperature is $T_d = 18 \pm
4\,$K. The pressure required to bound the clumps is 
$\log(k^{-1}\,P\ {\rm cm}^{3}\,{\rm K}^{-1}) = 6.1 \pm 0.3$. 
The scatter in $T_d$ and $k^{-1}\,P$ values is only marginally
larger than the uncertainty in any individual source measurement.  A
comparison with the results of our previous papers must take into
consideration the change in the opacity law and the inclusion of
additional turbulent support within each clump.  Together these
changes almost exactly cancel in the determination of the bounding
pressure and approximately halve the computed internal
temperature. With this in mind, the clump temperatures in Orion B
South appear similar to those in Orion B North (Paper III) and
somewhat warmer than those in Ophiuchus (Paper II). The bounding
pressure is lowest in Orion B and highest in Ophiuchus.  As
noted in Paper III, this is in general agreement with the known
conditions within the clouds; 
%
%
Ophiuchus has a significantly higher
{central column density, and hence internal pressure, 
than Orion, where observations have been 
obtained. In Orion the typical peak extinction 
through the cloud is 8 magnitudes with a maximum of  
17 magnitudes (Lambardi and Alves 2001), whereas in 
the core of Ophiuchus the peak extinction reaches 36 magnitudes
(Johnstone et al. 2004).}
Thus the Orion molecular
clouds are bathed in an enhanced interstellar radiation field which
warms the region.

\subsection{Clump Temperatures Derived from Submillimeter Colors}

In principle, the dust temperature within each clump can be estimated
from the submillimeter spectral index, $\gamma$, where $S(\nu) \propto
\nu^{\gamma}$. Assuming that the dust opacity follows a power-law with
$\kappa(\nu) \propto \nu^{\beta}$, and that the dust emission can be
approximated by a blackbody at temperature $T_d$, the dust emission is
given by
\begin{equation}
S(\nu) \propto \nu^{\beta}\,B_{\nu}(T_d).
\end{equation}
In practice, however, determining the dust temperature is complicated
by the fact that the dust opacity power-law coefficient, $\beta$, is
poorly constrained and often varies with location and physical
conditions (Goldsmith, Bergin, \& Lis 1997; Visser et al. 1998;
Hogerheijde \& Sandell 2000; Beuther, Schilke, \& Wyrowski 2004). As
well, the 450\,$\mu$m and 850\,$\mu$m beam sizes at the JCMT are
significantly different and the 450\,$\mu$m beam is not well
approximated by a single Gaussian function (Hogerheijde \& Sandell
2000); at least half of the flux in the 450\,$\mu$m beam is associated
with a broad component about $30\arcsec$ in extent.  Thus, simple
convolution of the central $8.5\arcsec$ 450\,$\mu$m beam with the
$14.5\arcsec$ 850\,$\mu$m beam will significantly overestimate the
contribution of 450\,$\mu$m emission on large scales. A sophisticated
solution to determining $\beta$ is to use a Fourier Transform Spectrometer
to produce a low resolution spectrum of the 850\,$\mu$m emission alone.
At present, however, this is only possible for very bright sources 
(Friesen et al. 2005). Here we utilize the 450\,$\mu$m and 850\,$\mu$m
emission from SCUBA and follow the approach of Reid \& Wilson (2005) in
preparing the maps for analysis. We convolve the 850\,$\mu$m
map with a model of the 450\,$\mu$m beam and the 450\,$\mu$m map with
a model of the 850\,$\mu$m beam. For this purpose the 450\,$\mu$m
beam is modeled as two Gaussian components with 8.5\arcsec and 30\arcsec\ FWHM,
and relative peak intensities of 0.95 and 0.05. The 850\,$\mu$m
beam is modeled as two Gaussian components with 14.5\arcsec and 30\arcsec\ FWHM,
and relative peak intensities of 0.95 and 0.05. 

\begin{figure}[htp]
\includegraphics[width=0.65\textwidth,angle=90,clip]{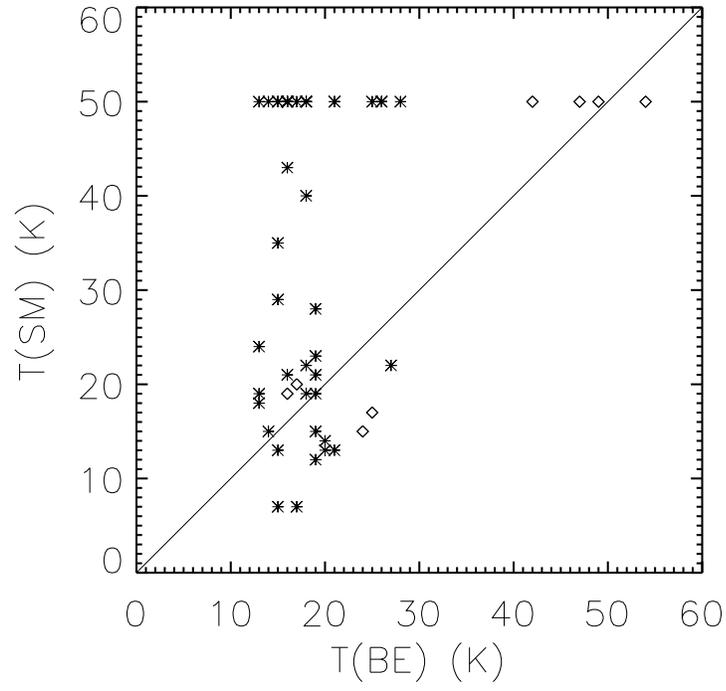}
\caption{\label{f_obs_temp}
Comparison of the clump temperature derived from the Bonnor-Ebert 
analysis versus the clump temperature derived from spectral 
energy fitting between the 450\,$\mu$m and 850\,$\mu$m observations
(see text). }
\end{figure}

{Assuming that $\beta = 2$ (Hildebrand 1983), 
the central dust temperature for each clump
was estimated using the total flux in the convolved 850\,$\mu$m and
450\,$\mu$m maps by inverting  
\begin{equation}
{S_{450} \over S_{850} } = 24\,\left[ exp\left( {17 \over T_d} \right) 
- 1 \over exp\left({32 \over T_d}\right) - 1\right].
\end{equation}} 
The derived values are provided in Table \ref{t_450850} and a
comparison with the derived Bonnor-Ebert model temperatures is shown
in Figure \ref{f_obs_temp}. It is clear from the figure that while
there is qualitative agreement on which clumps are warm, the spectral
energy distribution derived temperatures for most clumps are poorly
constrained by the available submillimeter-wavelength data.  Some of
the clumps, particularly at the faint end of the distribution, have
450\,$\mu$m to 850\,$\mu$m flux ratios which are inconsistent with
modified blackbodies having $\beta = 2$. In Table \ref{t_450850} and
Figure \ref{f_obs_temp} we assume that these sources are warm ($T_d =
50\,$K), although alternative explanations include $\beta > 2$ and/or
measurement unreliability for the derived fluxes.  {We
have already noted in Section~\ref{s_odr} that the typical
uncertainty in the flux determination at 850\,$\mu$m is 20\% whereas
at 450\,$\mu$m the typical uncertainty is 50\% (Jenness et al. 2002),
notwithstanding the complication of convolution discussed
above.} Setting aside the atypical high flux ratio
sources, the mean dust temperature of the clumps derived from fitting
the spectral energy distribution between 450\,$\mu$m and 850\,$\mu$m
is $T_d = 18 \pm 6\,$K.

Once protostellar source catalogues obtained from Spitzer Space Telescope
observations become available for the Orion B South region, a detailed
investigation between the observed concentration and temperatures of
clumps derived from the continuum observations and the presence of
internal sources will be possible, yielding significant improvement in 
our understanding of the evolution of these prestellar and protostellar 
envelopes.

\subsection{Clump Temperatures Derived from Molecular Line Observations}

\begin{figure}[htp]
\includegraphics[width=0.5\textwidth,angle=-90,clip]{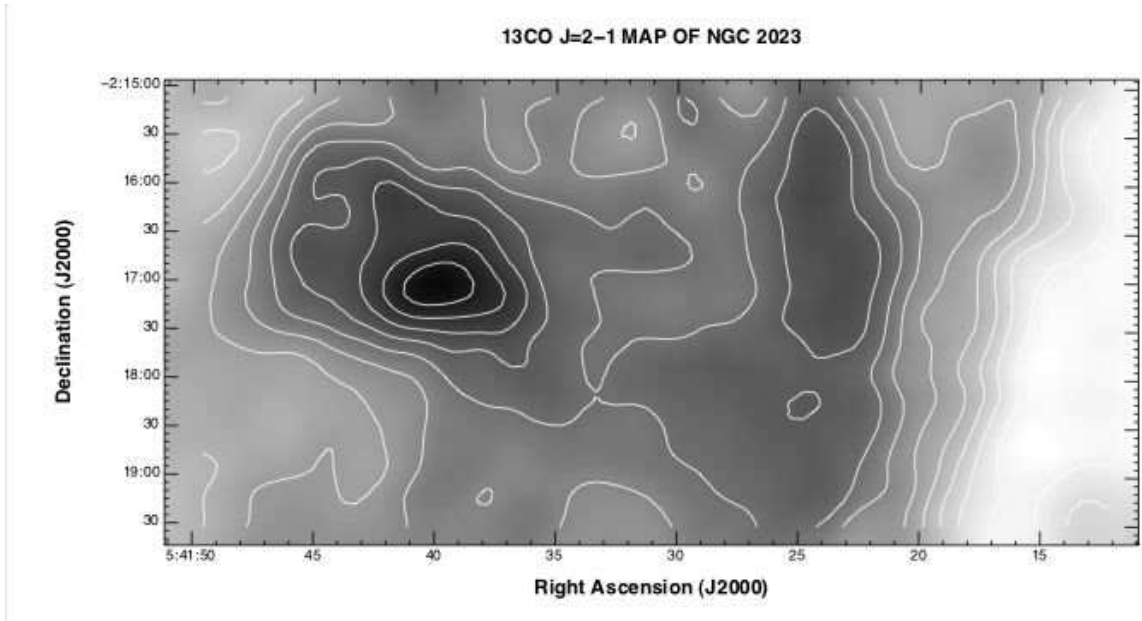}
\caption{\label{f_13co21}
A $^{13}$CO J = 2--1 raster map of NGC 2023. The beam size is $21\arcsec$. 
{The sampling interval is $7.5\arcsec$ in R.A.\ and Dec. The maximum 
contour is 58.0 K.km/s and the contour interval is 5.0 K.km/s.}}
\end{figure}

\begin{figure}[htp]
\includegraphics[width=0.5\textwidth,angle=-90,clip]{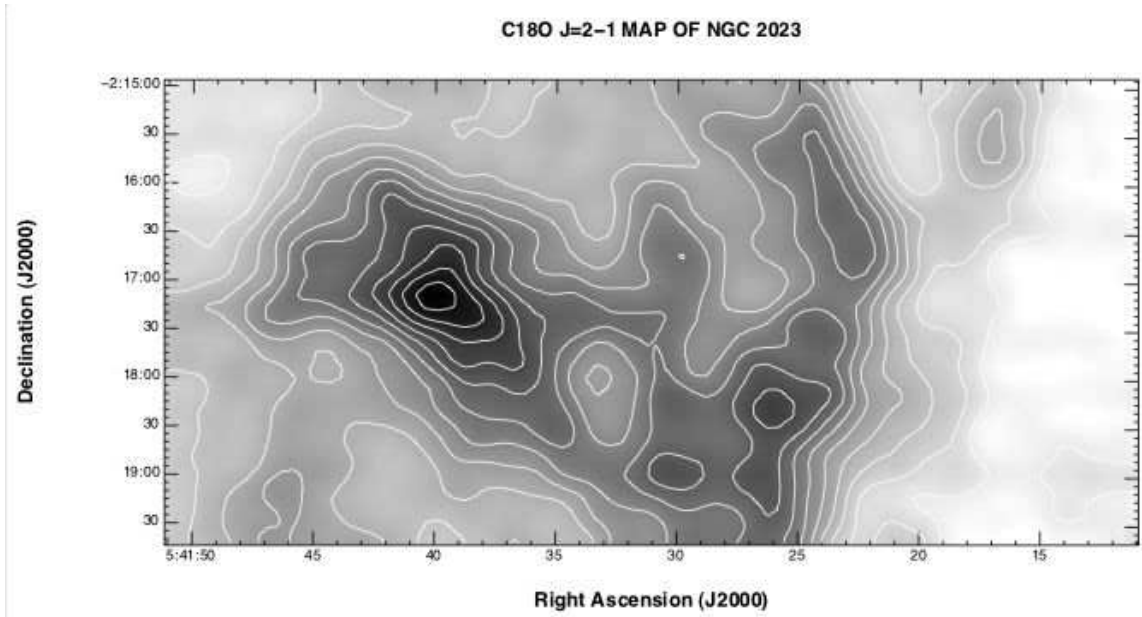}
\caption{\label{f_c18o21}
A C$^{18}$O J = 2--1 raster map of NGC 2023. The beam size is $21\arcsec$. 
{The sampling interval is $7.5\arcsec$ in R.A.\ and Dec. The $^{13}$CO J = 
2--1 and C$^{18}$O J = 2--1 maps were obtained simultaneously with a single 
tuning of the facility A-band receiver.} The maximum contour is 12.3 K.km/s 
and the contour interval is 1.0 K.km/s}
\end{figure}

\begin{figure}[htp]
\includegraphics[width=0.5\textwidth,angle=-90,clip]{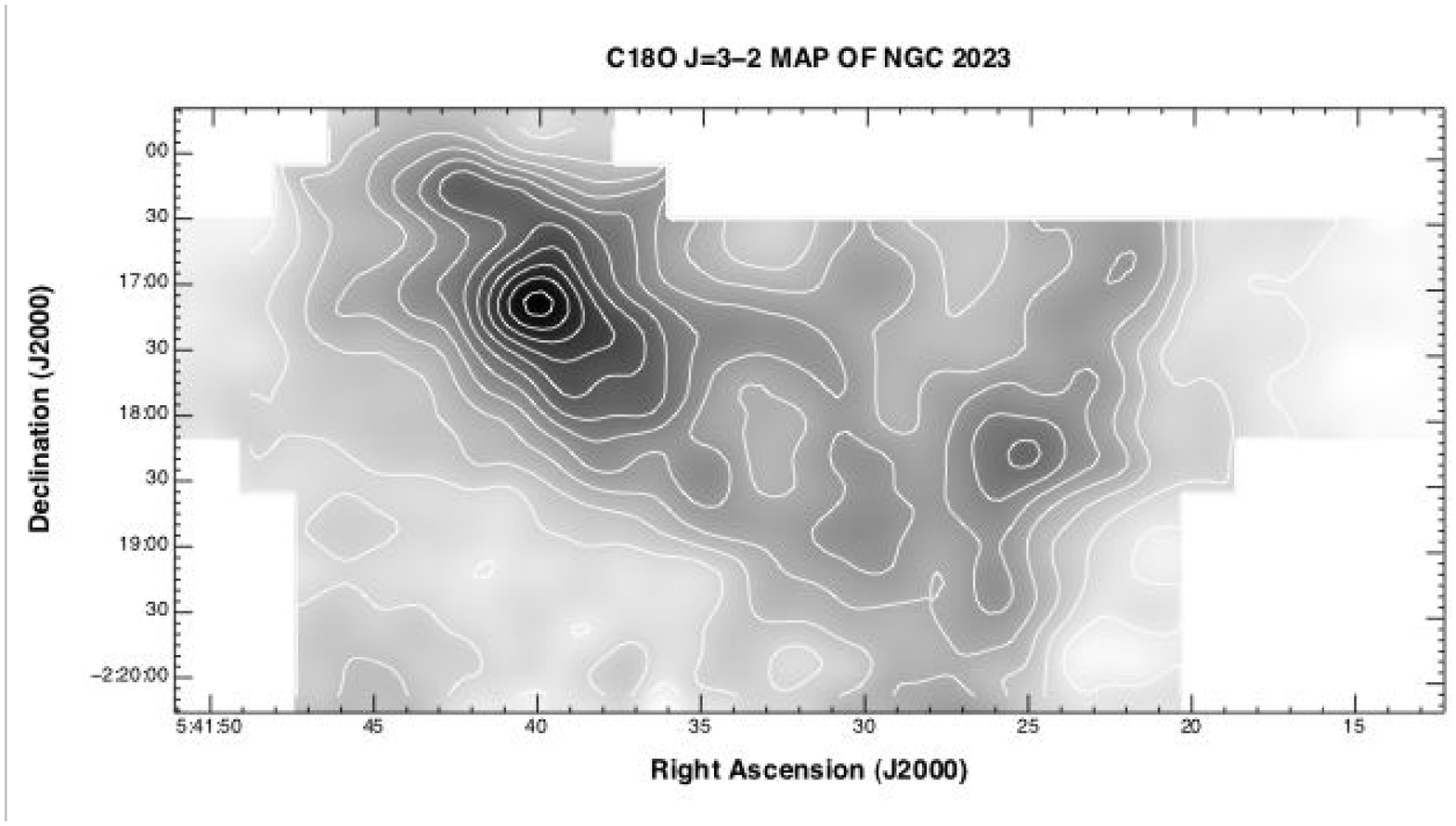}
\caption{\label{f_c18o32}
A C$^{18}$O J = 3--2 raster map of NGC 2023. The beam size is $14\arcsec$. 
{The sampling interval is $5\arcsec$ in R.A.\ and Dec. The maximum contour
is 13.4 K.km/s and the contour interval is 1.0 K.km/s.} }
\end{figure}

\begin{figure}[htp]
\includegraphics[width=0.5\textwidth,angle=-90,clip]{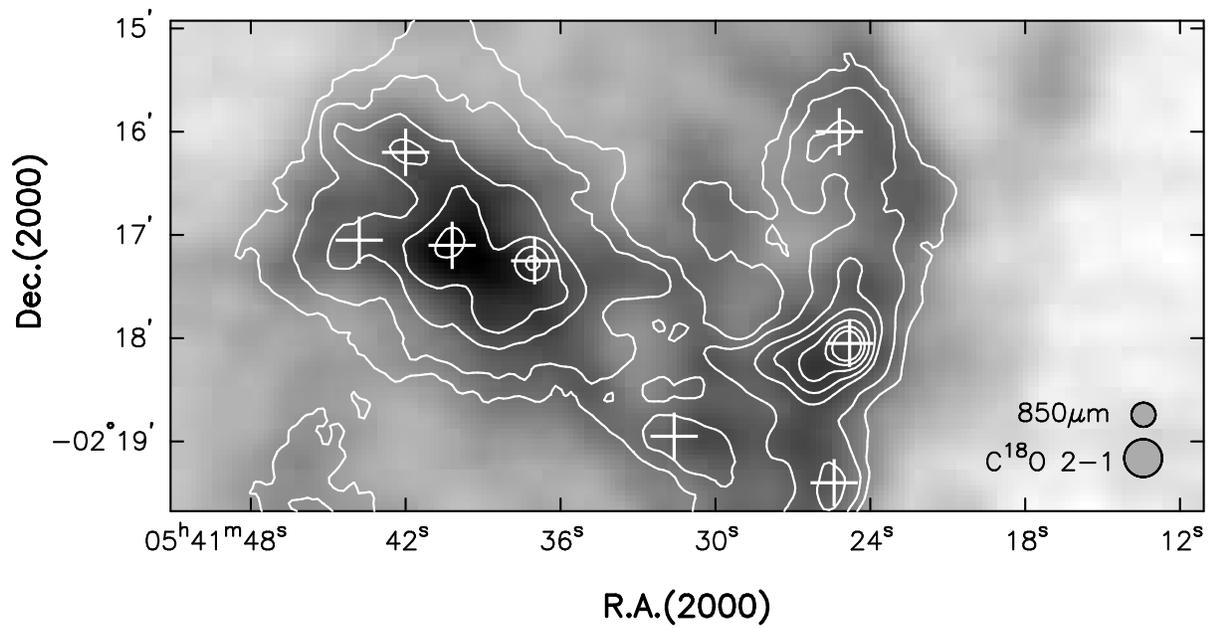}
\caption{\label{f_overlay}
The C$^{18}$O J = 2--1 grayscale image of NGC 2023 (see Figure 12), overlain
by contours of the  850\,$\mu$m continuum emision. The contour levels shown
are 0.1, 0.2, 0.4, 0.6, 1.0, 1.5, and 2.0 Jy/beam, and the beamsizes to 
half-power are indicated at lower right. The superposed crosses show the 
locations of the continuum sources indicated by the automated 
clump-finding routine (see Section 3.1, and also Figures 3 and 5). }
\end{figure}

Mapping of CO isotopomers was carried out at the JCMT during the same 
observing runs, when sky conditions did not permit continuum observing.  
Scan maps of part of the Orion B South region, NGC\,2023, 
were obtained in $^{13}$CO J = 2--1, C$^{18}$O J = 2--1, 
and C$^{18}$O J = 3--2. The maps are shown in Figures \ref{f_13co21}, 
\ref{f_c18o21}, and \ref{f_c18o32}. Figure \ref{f_overlay} shows 850\,$\mu$m 
continuum contours overlaid on the C$^{18}$O J = 2--1 map. The J = 2--1 maps 
were  acquired using $7.5\arcsec$ sampling, which is about one 
third of the $21\arcsec$ beamwidth. The C$^{18}$O J = 3--2 map has a 
beamwidth of $14\arcsec$ and was acquired using $5\arcsec$ sampling.

The $^{13}$CO to C$^{18}$O J = 2--1 line ratio provides the optical depth 
over the map, on the assumption of LTE. {Goldsmith, Bergin, and Lis (1997) 
find that the abundance ratio, $^{13}$CO/C$^{18}$O, lies in the range 6-15. In 
the present
analysis we assume that the abundance ratio of $^{13}$CO to C$^{18}$O 
is 10}.  As expected, the C$^{18}$O 
emission is optically thin, with $\tau$ being typically less than 0.1,   
and ranging up to a maximum of 0.2. In addition to being optically 
thin, the C$^{18}$O J = 3--2 map has the same resolution as the 850\,$\mu$m
map and so is the most useful for a comparison of dust and gas emission. 
The C$^{18}$O and continuum maps agree quite well in general appearance, but  
a closer examination shows that the C$^{18}$O emission is somewhat 
smoother, less clumpy, and more extended than the dust continuum emission. 
These differences might be due in part to the removal of extended continuum 
emission by the SCUBA chopping technique, in part to different dependence of 
CO and dust emission on temperature, and in part to the freeze-out of CO 
molecules onto grains. 

We have obtained the CO excitation temperature from the ratio of the two 
rotational lines of C$^{18}$O. The relevant equation is 
\begin{equation}
T_{\rm ex} = 16.59\,\log_{e}\left[{4\,I_R(3-2) \over 9\,I_R(2-1)}\right].  
\end{equation}
{This equation is obtained from standard LTE and optically thin
expressions and is derived in Mitchell (1993). Here, $I_R$ is the
integrated intensity of the appropriate spectral line, with the line
temperature expressed in units of radiation temperature ($T_R$ =
$T_A^*$/$\eta_{mb}$, with the assumption that the beam filling factor
is unity).  We use appropriate beam efficiencies for the two
transitions ($\eta_{mb}$ = 0.69 for the 2-1 line and 0.63 for the 3-2
line) and convolve the C$^{18}$O J = 3--2 observations to match the
$21\arcsec$ beam of the J = 2--1 data.} We find that, for most pixels
over the map, the C$^{18}$O excitation temperature lies in the range
14\,K to 20\,K, with a mean of 16\,K.  The (rare) extreme values are
10\,K and 25\,K. The uncertainty in individual values, due only to
noise in the data, is $\pm 2\,$K.  These excitation temperatures are
quite consistent with the average temperature of $18 \pm 4\,$K found
above for clumps in the NGC\,2023 region, using a Bonnor-Ebert
analysis.

Five of the continuum clumps in Table \ref{t_850data} 
fall inside the region of NGC\,2023 
that we mapped in all three CO transitions. {Using the continuum positions 
of these clumps, we have calculated the C$^{18}$O excitation temperature 
from the above equation and the 
C$^{18}$O column density using the expression given by Mitchell et al. (2001).}
Optically thin values of column density were multiplied by the factor 
$\tau/(1-\exp(-\tau))$ to correct for optical depth.  The C$^{18}$O column 
density was converted to a H$_2$ column density using the ratio 
N(H$_2$)/N(C$^{18}$O) $ = 6 \times 10^6$.  In Table \ref{t_2023} we list, for 
these five continuum clumps, the temperature from the Bonnor-Ebert analysis, 
the dust temperature from the 850\,$\mu$m/450\,$\mu$m  ratio, the C$^{18}$O 
excitation temperature, the H$_2$ column density derived from the 850\,$\mu$m 
peak flux per beam (assuming the BE derived temperature), 
the H$_2$ column density from the C$^{18}$O observations, 
and the ratio of the two column density  determinations. 

The gas temperatures agree very well with the dust temperatures. This
close agreement is rather unexpected in view of the large
uncertainties in the two dust temperature techniques.  The single
exception is 054153-02189, where the dust temperature from the
submillimeter continuum ratio is anomalously low.  Column densities
derived from CO and dust emission agree reasonably well for four of
the clumps (054153-02189, 054162-02172, 054167-02171, 054173-02170)
but differ by a large factor for one (054141-02180).  A similar
analysis for 16 clumps in Orion B North (Mitchell et al. 2001)
resulted in masses from dust continuum exceeding masses from C$^{18}$O
by factors of a few. The present closer agreement is due in part to
the use of a larger grain opacity in this paper.

Freeze-out of CO molecules on grains is a common phenomenon in cold
cores (e.g. J\"orgensen, Sch\"oier, and van Dishoeck 2005) and could
explain the high N(850)/N(C$^{18}$O) ratio for the one clump. However,
since all five clumps have similar temperatures, why has molecular
depletion acted in only one of the five clumps? Various uncertainties
enter the column density determinations, including the value of the
grain opacity and CO isotopic abundances, so that the ratio,
N(850)/N(CO), could easily be increased by a factor of two or three
and become consistent with some CO freeze-out in all cases.  An
examination of the C$^{18}$O J = 3--2 map shows that obvious clumps
are present at only two of the five positions, namely continuum clumps
054141-02180 and 054167-02171. The apparent agreement in the case of
the first clump is deceptive, however. The C$^{18}$O peak is displaced
$20\arcsec$ to the southeast of the continuum peak. Although there is
an extension of 850\,$\mu$m emission from this clump towards the
southeast, the clump-finding algorithm did not detect an independent
clump at this position.  Examination of the higher resolution
450\,$\mu$m map does show a separate peak at the C$^{18}$O
position. In fact, this second peak was detected by Sandell et
al. (1999) in an early SCUBA observation. They called this clump
054141-02180 NGC 2023 mm1 and they named the source to the southeast
NGC 2023 mm2. The position of the latter source 
%
%
coincides with the C$^{18}$O peak. From a 
variety of evidence, including a CO bipolar outflow, Sandell et al. (1999) 
suggested that NGC 2023 mm1 is a Class 0 YSO. It appears that strong CO 
depletion has occurred in mm1 but not in mm2. Clump 054167-02171 is bright 
in both continuum emission and in C$^{18}$O emission. It is almost 
certainly significant that this clump is the warmest of the five clumps, 
with a temperature of 24\,K. This is warm enough that CO depletion is not 
expected to occur (e.g. Figure 1 of J\"orgensen et al. 2005).

\subsection{The Clump Mass Distribution Function}

Stars form inside dense condensations within Giant Molecular Clouds.
At least some of the clumps investigated above are expected to form,
or to be forming, stars, and thus understanding the physical properties
of these clumps and their environment will help constrain formation
mechanisms.  One important bulk property of the clumps is the
distribution of clumped mass.  Following the pioneering study of the 
Ophiuchus molecular cloud complex 
by Motte et al. (1998), a number of submillimeter-wavelength
surveys have been completed, each resulting in essentially the same
conclusion, despite 
%
%
the objects found being in different environments.  
The mass
function of the submillimeter condensations (Testi \& Sargent 1998;
Paper II; Paper III; Motte et al. 2001) essentially mimics the stellar
IMF where $N(M) \propto M^{-1.35}$ for stars more massive than $M \sim
0.5\,M_\odot$ (Salpeter 1955) and the bulk of the mass resides at the 
low-mass end
(Scalo 1986; Kroupa et al.  1993).  This result is in direct contrast
with the mass function of molecular clouds (Williams \& McKee 1997;
McKee \& Williams 1997; Kramer et al. 1998; Williams, Blitz, \& McKee
2000 and references therein) where $N(M) \propto M^{-0.5}$ and the
majority of the mass is found in the few large clouds.

\begin{figure}[htp]
\includegraphics[width=0.65\textwidth,angle=90,clip]{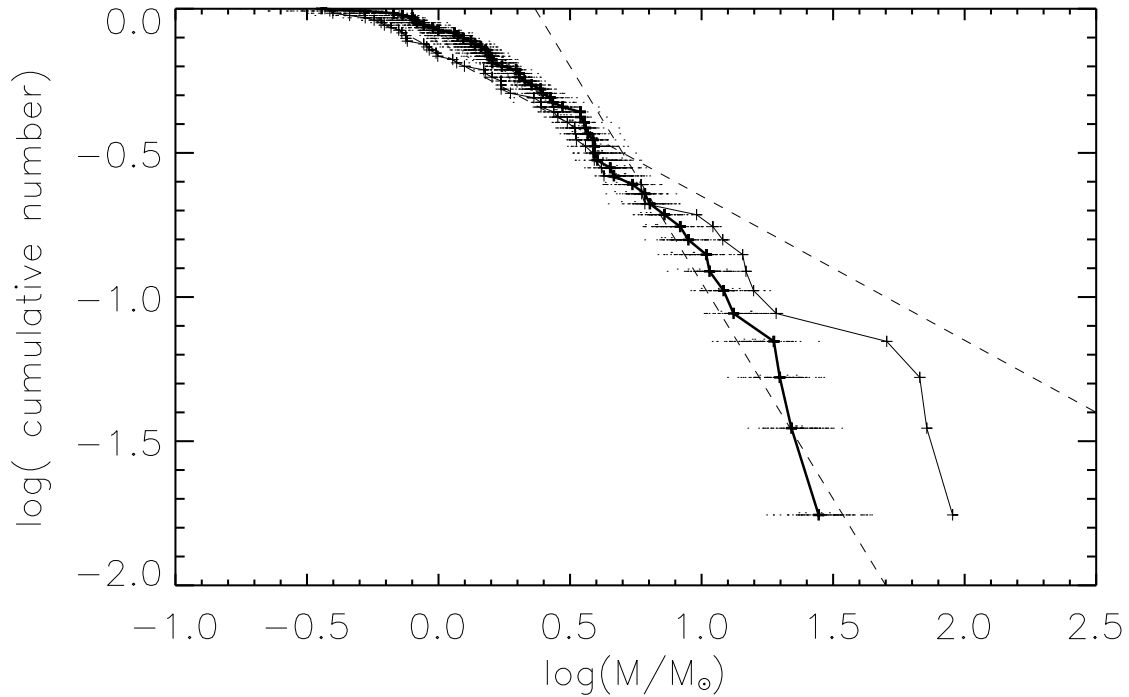}
\caption{\label{f_obs_cum} 
Cumulative number function for the 850\,$\mu$m clumps in Orion B South.
The thin line converts the measured flux directly to mass using a
constant temperature $T_d = 20\,$K. The thick line converts the flux
to mass using the derived temperature from fitting the clumps to
Bonnor-Ebert spheres (see text).  The horizontal lines represent the
extent to which the masses might change due to uncertainties in the
Bonnor-Ebert Measurements {(see Paper III)}. The steep dashed line 
has a slope $M^{-1.5}$ and approximates the high-mass end of the 
{Bonnor-Ebert} cumulative distribution.  The shallow dashed line 
has a slope $M^{-0.5}$ and
approximates the low-mass end of the distribution. The low-mass end
is likely to be severely incomplete (see text).}
\end{figure}

As in Papers II and III, the clump mass distribution is plotted in
terms of cumulative number $N(M)$ of clumps with masses greater than
$M$.  Figure \ref{f_obs_cum} presents the Orion B South data for both
a constant flux-to-mass ratio (for $T_d = 20\,K$) and
for masses calculated using the Bonnor-Ebert sphere analysis.  The
horizontal dotted lines indicate the possible variation of the
distribution due to uncertainties in the Bonnor-Ebert derived
temperatures (as discussed in Paper III). The two dashed lines in
Figure \ref{f_obs_cum} denote power-law fits with $N(M) \propto
M^{-\alpha}$, where $\alpha = 0.5$ (shallow line) or $\alpha = 1.5$
(steep line) {and are presented to guide the eye}.  The cumulative
mass function determined using Bonnor-Ebert derived masses is steep,
as is the stellar IMF at the high mass end, but flattens significantly
around $M \sim$ 3 -- 10\,$M_\odot$. This is in contrast to the clump
IMF measured in Orion B North (Paper III) and Ophiuchus (Paper II)
using the same technique {and with similar sensitivity}.  In both
those distributions the steep mass function continued to $M \sim 1
M_\odot$.

One explanation for the observed discrepancy may be the environmental
conditions in which the clumps are found.  The typical clump in Orion
B South is more extended, and thus less dense, than those found in
Orion B North or Ophiuchus.  The lower density of the clumps may imply
a lower bounding pressure, in agreement with the Bonnor-Ebert analysis
above.  Why the clump mass distribution should be strongly affected by
a relatively small change in bounding pressure is, however,
unclear. The Jeans' mass at which clumps become unstable varies only
with the square root of the bounding pressure. More likely, the effect
is due to the sensitivity limits of the survey. The mass versus radius
distribution of clumps in Figure \ref{f_obs_mr} reveals that
incompleteness in clump identification may already become important at
$M \sim 5\,M_\odot$, explicitly because the clumps are more diffuse,
lower in surface brightness, and thus harder to detect. 

There is significant discrepancy between the mass function of
structure in molecular clouds derived from molecular line observations
(shallow), particularly CO (e.g. Kramer et al. 1998), and
submillimeter-wavelength dust continuum radiation (steep). 
It is important to
note that there are exceptions to this trend. Onishi et al. (2002)
identified clumps within Taurus using H$^{13}$CO$^{+}$ J = 1--0 line 
emission and
found a steep mass function. Also, Kerton et al. (2001) and Mookerjea
et al. (2004) have found much shallower mass functions associated with
larger, more massive dust continuum clumps in distant Giant Molecular
Clouds.  These latter submillimeter-wavelength 
structures are most definitely the counterparts
to the dense cores typically observed in molecular line emission, while
the H$^{13}$CO$^{+}$ clumps are counterparts to the small-scale, dense dust
continuum structures discussed in this paper. It remains to be
determined, however, whether there is a clear separation between these
two types of structure or whether the different observing techniques
are presently sensitive only at the extremes of an underlying smoothly 
varying distribution.

\subsection{Clump Environment}

\begin{figure}[htp]
\centering
\includegraphics[width=0.8\textwidth,clip]{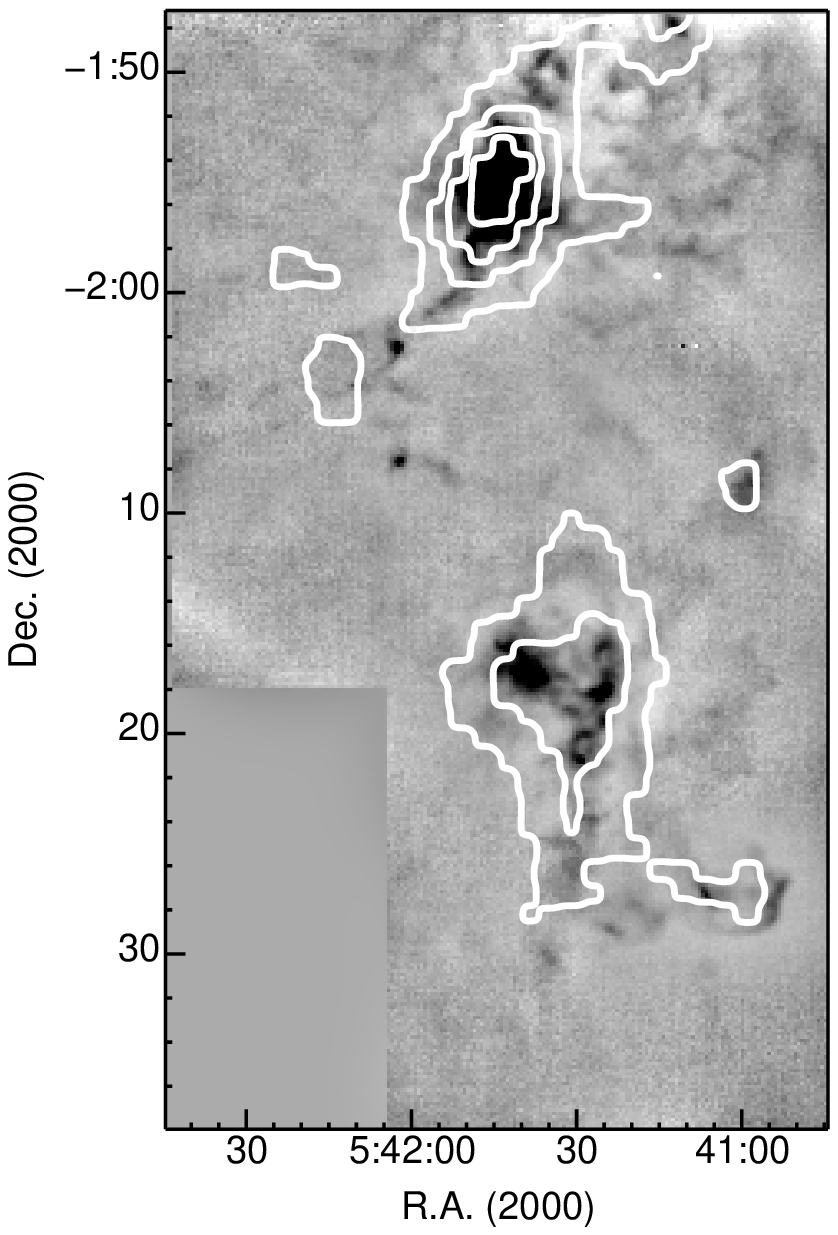}\\
\caption{Contours of CS J = 2--1 emission (from Lada et al. 1991) are
overlaid on the 850\,$\mu$m map. The contour levels are at 1.5, 3.0, 6.0,
and 12.0 K\,km\,s$^{-1}$. Note that a majority of the strong dust continuum
sources lie above the 3 K\,km\,s$^{-1}$ contour.
}\label{f_cs}
\end{figure}

The cold dust clumps found are grouped in three main
concentrations: B33, NGC\,2024, and NGC\,2023.
%
%
These regions were also observed to be bright 
in line emission; for example, in addition to CO and 
its isotopomers, CS J = 2--1, traditionally a dense gas tracer (LBS).  
Figure \ref{f_cs} overlays the integrated CS 
observations on the 850\,$\mu$m image for comparison. As in Orion B North
(Paper III), the vast majority of the submillimeter-wavelength clumps 
lie above
the 3\,km\,s$^{-1}$ contour, corresponding to column densities of
$N_H \sim 10^{22}\,$cm$^{-2}$ through the CS core region. This
is consistent with the result that an extinction threshold may exist
for the formation of smallscale, thermally dominated, submillimeter
clumps (Paper II, Paper III, Johnstone, Di Francesco, \& Kirk 2005).
There are, however, two distinct clumps (SMM J054204-02077 and SMM 
J 054205-02025) residing in particularly low integrated CS emission.
Both of these sources are highly concentrated but otherwise unremarkable.

Johnstone et al.  (2005) noted that approximately twenty percent of the
mass of the core region in Ophiuchus has fragmented into identifiable 
submillimeter clumps. The apparent threshold for submillimeter clump 
formation, however, leads to few sources being detected outside the dense 
core and thus only two percent of the mass of the entire Ophiuchus cloud 
is visible as {submillimeter clump} fragments. 
These are intriguing numbers, comparable to the
fractional mass of embedded young clusters in dense cores (Lada \& Lada
2003), suggesting that the clumping process may be linked to the formation 
of stellar associations. Comparing the CS mass determinations of the parent 
cores, derived by LBS, against the total submillimeter clump mass residing 
within each core for both Orion B South and Orion B North (Paper III) is 
straightforward. Although there exists a significant spread from 5\% mass 
fragmentation in NGC 2071 to  35\% mass fragmentation in NGC 2024, 
overall one finds $\sim 300\,M_\odot$ in submillimeter clumps within
$\sim 1700\,M_\odot$ of CS clumped material, or approximately 20 percent
fragmentation by mass. 
This result is identical to that found for the Ophiuchus region.

\section{Conclusions}\label{s_sum}

We have presented images of the southern part of the
Orion B region obtained at submillimeter wavelengths of 850 and
450\,$\mu$m, covering an area 1300 arcmin$^2$ in extent.  Utilizing the
techniques developed in Papers I, II, and III, we have identified 57
independent dust concentrations, or ``clumps'', within this region. The
majority of these clumps are concentrated in two star-forming regions,
NGC\,2024 and NGC\,2023; a third minor grouping appears in the
Horsehead Nebula cloud. Most of the dust mass is concentrated in
NGC\,2024.

We have been able to derive estimated masses and temperatures for the
majority of the clumps by noting that they can be modelled as
Bonnor-Ebert spheres; that is, they can be approximated by almost
constant density models with low internal velocities. On this basis
the typical clump temperature is found to be 18$\pm$4~K; this value is
in reasonable agreement with that (21$\pm$9~K) derived from the
spectral indices obtained from the 850 and 450\,$\mu$m data.  Thus for
most of the clumps a temperature of 20~K is a good working
approximation.  However, 9 clumps are centrally concentrated enough
that they may be collapsing, and 4 of these objects are particularly
bright at 850\,$\mu$m. All of the latter lie within
NGC\,2024 and have temperatures which are considerably
higher than that for the majority of the clumps.

Using CO isotopomer spectral line data obtained for NGC2023 we have
attempted to obtain independent estimates of temperatures and masses
for the clumps in this particular region. With one exception the
results obtained agree with those derived from the continuum
data. It can be difficult, however, to clearly relate CO structures
with equivalent continuum clumps. This problem becomes more acute at
the lower mass end of the clump distribution, where continuum
sensitivity is limited and individual clumps overlap. 

The mass function of submillimeter clumps found in Orion B South is
steep, in agreement with the results in other nearby regions
(Motte et al. 1998; Paper II; Paper III; Motte et al. 2001).
The fraction of mass in each CS-identified core which has fragmented
in observable submillimeter clumps is $\sim 20$ percent. 

\acknowledgments

The research of D.J.\ and G.M.\ is supported through grants from the
Natural Sciences and Engineering Research Council of Canada.  Thanks
to John Bally for supplying the CS data and lengthy conversations.
Thanks also to Rachel Friesen and Helen Kirk for critical readings and
to the anonymous referee for comments that improved this paper. We wish
to acknowledge our collaborators in this JCMT key project survey of
molecular clouds: Lorne Avery, Shantanu Basu, Mike Fich, Jason Fiege,
Gilles Joncas, Lewis Knee, Brenda Matthews, Ralph Pudritz, Gerald
Schieven, and Christine Wilson.  We thank Tim Jenness and Dave Berry for help
with converting one of the figures. The JCMT is operated by the Joint
Astronomy Centre on behalf of the Particle Physics and Astronomy
Research Council of the UK, the Netherlands Organization for
Scientific Research, and the National Research Council of Canada.  The
authors acknowledge the data analysis facilities provided by the
Starlink Project which is run by CCLRC on behalf of PPARC. The
Digitized Sky Survey was produced at the Space Telescope Science
Institute under U.S. Government grant NAG W-2166.  We have also made
use of data products from the Midcourse Space Experiment (MSX).
Original processing of the latter data was funded by the Ballistic
Missile Defense Organization with additional support from NASA Office
of Space Science. Guest User, Canadian Astronomy Data Centre, which is 
operated by the Dominion Astrophysical Observatory for the National 
Research Council of Canada's Herzberg Institute of Astrophysics.





\begin{deluxetable}{lccrrclrcrr}
\tabletypesize{\scriptsize}
\tablewidth{480pt}
\tablecaption{Clump properties in Orion B South derived from  850\,$\mu$m data.\label{t_850data}}
\tablehead{
\colhead{Name\tablenotemark{a}}&
\colhead{R.A.\tablenotemark{b}}&
\colhead{Dec.\tablenotemark{b}}&
\colhead{$S_{850}$\tablenotemark{c}}&
\colhead{$S_{850}^{\rm peak}$\tablenotemark{c}}&
\colhead{$R_{\rm eff}$\tablenotemark{c}}&
\colhead{$C$\tablenotemark{d}}&
\colhead{$T_d$\tablenotemark{d}}&
\colhead{log$P/k$\tablenotemark{d}}&
\colhead{$M_{T_d}$\tablenotemark{d}}&
\colhead{$M_{T_d=20\,{\rm K}}$\tablenotemark{e}}\\
\colhead{(SMM J)}&
\colhead{(J2000)}&
\colhead{(J2000)}&
\colhead{(Jy)}&
\colhead{(Jy/bm)}&
\colhead{(10$^3$ AU)}&
\colhead{}&
\colhead{(K)}&
\colhead{(K/cm$^{3}$)}&
\colhead{($M_\odot$)}&
\colhead{($M_\odot$)}
}

\startdata
054086-01528&  05:40:51.8& -01:52:48&  1.23&  0.18&   11&   0.40& 18&    6.0&   1.15&   0.98\\
054089-02271&  05:40:53.4& -02:27:06&  2.18&  0.31&   12&   0.51& 16&    6.0&   2.44&   1.73\\
054090-01531&  05:40:54.0& -01:53:06&  0.89&  0.21&    9&   0.42& 16&    6.2&   1.00&   0.71\\
054091-02278&  05:40:54.6& -02:27:51&  2.90&  0.28&   14&   0.45& 19&    6.0&   2.49&   2.31\\
054096-02074&  05:40:57.6& -02:07:27&  2.18&  0.28&   13&   0.54& 15&    5.9&   2.72&   1.73\\
054097-02086&  05:40:58.2& -02:08:36&  4.21&  0.40&   16&   0.57& 18&    5.9&   3.92&   3.34\\
054100-02092&  05:40:59.8& -02:09:15&  5.24&  0.29&   19&   0.52& 19&    5.8&   4.49&   4.16\\
054105-01558&  05:41:03.2& -01:55:51&  1.43&  0.20&   11&   0.45& 16&    6.0&   1.60&   1.13\\
054112-02273&  05:41:07.0& -02:27:18&  3.08&  0.48&   13&   0.61& 16&    6.0&   3.46&   2.45\\
054116-01580&  05:41:09.4& -01:58:00&  1.25&  0.24&   10&   0.48& 15&    6.1&   1.56&   0.99\\
054140-01507&  05:41:23.8& -01:50:45&  0.96&  0.27&    9&   0.52& 13&    6.1&   1.53&   0.76\\
054141-02180&  05:41:24.8& -02:18:03& 18.57&  3.19&   21&   0.87& 25&    5.9&  10.72&  14.74\\
054141-02289&  05:41:24.4& -02:28:54&  0.76&  0.21&    8&   0.44& 15&    6.2&   0.95&   0.60\\
054142-02160&  05:41:25.2& -02:16:00&  7.42&  0.47&   20&   0.60& 19&    5.8&   6.36&   5.89\\
054142-02194&  05:41:25.4& -02:19:24&  4.18&  0.48&   14&   0.55& 19&    6.1&   3.58&   3.32\\
054146-02208&  05:41:27.8& -02:20:51&  2.36&  0.29&   12&   0.45& 19&    6.1&   2.03&   1.88\\
054146-02239&  05:41:27.6& -02:23:54&  1.90&  0.26&   12&   0.50& 16&    6.0&   2.13&   1.51\\
054148-02200&  05:41:28.6& -02:20:03&  4.92&  0.48&   15&   0.54& 20&    6.1&   3.91&   3.91\\
054149-02212&  05:41:29.4& -02:21:15&  5.36&  0.97&   15&   0.77& 17&    5.8&   5.46&   4.26\\
054149-02232&  05:41:29.2& -02:23:15&  0.83&  0.25&    8&   0.50& 13&    6.2&   1.33&   0.66\\
054151-02251&  05:41:30.4& -02:25:06&  0.91&  0.19&    9&   0.38& 21&    6.1&   0.68&   0.73\\
054152-02262&  05:41:31.0& -02:26:12&  3.45&  0.21&   17&   0.45& 19&    5.8&   2.96&   2.74\\
054153-02189&  05:41:31.6& -02:18:57&  4.56&  0.32&   18&   0.56& 17&    5.8&   4.64&   3.62\\
054155-01497&  05:41:33.2& -01:49:45&  1.87&  0.32&   10&   0.46& 18&    6.2&   1.74&   1.49\\
054156-02209&  05:41:33.4& -02:20:54&  1.16&  0.25&    9&   0.47& 15&    6.1&   1.45&   0.92\\
054156-02250&  05:41:33.6& -02:25:03&  0.78&  0.18&    9&   0.42& 16&    6.1&   0.88&   0.62\\
054158-01558&  05:41:34.8& -01:55:48&  4.15&  0.40&   15&   0.54& 18&    6.0&   3.87&   3.30\\
054158-02303&  05:41:34.6& -02:30:18&  0.47&  0.19&    6&   0.39& 16&    6.3&   0.52&   0.37\\
054159-01514&  05:41:35.6& -01:51:27&  2.01&  0.34&   11&   0.51& 16&    6.1&   2.26&   1.60\\
054160-01492&  05:41:35.8& -01:49:12&  1.14&  0.41&    8&   0.54& 14&    6.3&   1.60&   0.90\\
054160-01564&  05:41:35.8& -01:56:24& 15.16&  0.58&   24&   0.53& 26&    5.9&   8.29&  12.03\\
054160-02300&  05:41:36.0& -02:30:00&  0.49&  0.18&    7&   0.36& 21&    6.3&   0.36&   0.39\\
054161-01495&  05:41:36.6& -01:49:30&  1.58&  0.44&    9&   0.54& 15&    6.3&   1.97&   1.25\\
054162-02172&  05:41:37.0& -02:17:15& 19.86&  1.66&   26&   0.82& 24&    5.7&  12.13&  15.76\\
054163-01509&  05:41:37.8& -01:50:54&  2.18&  0.34&   11&   0.46& 18&    6.2&   2.03&   1.73\\
054163-01558&  05:41:37.6& -01:55:48&  7.46&  0.58&   15&   0.44& 28&    6.2&   3.69&   5.92\\
054164-01504&  05:41:38.4& -01:50:24&  1.48&  0.28&   10&   0.44& 18&    6.2&   1.38&   1.17\\
054167-02171&  05:41:40.2& -02:17:06& 13.91&  1.12&   17&   0.59& 27&    6.2&   7.23&  11.04\\
054170-02162&  05:41:42.0& -02:16:12& 12.06&  0.68&   23&   0.65& 21&    5.7&   8.91&   9.57\\
054172-01543&  05:41:43.0& -01:54:21& 85.03& 11.11&   24&   0.86& 47&    6.4&  21.96&  67.50\\
054173-01527&  05:41:44.0& -01:52:45& 18.04&  0.85&   24&   0.64& 25&    5.8&  10.41&  14.32\\
054173-01547&  05:41:44.0& -01:54:42& 63.67&  7.99&   23&   0.85& 42&    6.3&  18.82&  50.54\\
054173-02170&  05:41:43.8& -02:17:03&  7.67&  0.49&   20&   0.60& 20&    5.8&   6.09&   6.09\\
054174-01556&  05:41:44.4& -01:55:39& 90.41& 15.35&   19&   0.83& 54&    6.7&  19.82&  71.77\\
054175-01560&  05:41:45.0& -01:56:00&113.37& 10.15&   29&   0.87& 49&    6.2&  27.86&  90.00\\
054182-01576&  05:41:49.4& -01:57:39& 24.19&  1.07&   27&   0.69& 26&    5.7&  13.23&  19.21\\
054182-01596&  05:41:49.2& -01:59:36&  3.58&  0.47&   14&   0.63& 16&    5.9&   4.02&   2.84\\
054185-01534&  05:41:51.2& -01:53:27&  0.66&  0.19&    8&   0.42& 15&    6.2&   0.83&   0.53\\
054188-02002&  05:41:53.0& -02:00:12&  1.11&  0.30&    8&   0.45& 16&    6.3&   1.24&   0.88\\
054190-02004&  05:41:54.0& -02:00:24&  0.95&  0.31&    8&   0.48& 15&    6.3&   1.18&   0.75\\
054192-01565&  05:41:55.0& -01:56:33&  4.86&  0.22&   19&   0.40& 26&    5.8&   2.66&   3.86\\
054195-02009&  05:41:56.8& -02:00:54&  0.50&  0.25&    6&   0.48& 13&    6.4&   0.80&   0.40\\
054204-02077&  05:42:02.4& -02:07:42&  3.09&  0.95&   12&   0.76& 16&    6.0&   3.47&   2.45\\
054205-02025&  05:42:02.8& -02:02:33&  3.89&  1.14&   11&   0.72& 18&    6.2&   3.62&   3.09\\
054207-02017&  05:42:04.2& -02:01:42&  0.72&  0.19&    8&   0.40& 17&    6.2&   0.73&   0.57\\
054207-02032&  05:42:04.0& -02:03:12&  0.58&  0.21&    7&   0.44& 14&    6.3&   0.81&   0.46\\
054218-02042&  05:42:10.8& -02:04:12&  0.95&  0.30&    9&   0.56& 13&    6.2&   1.51&   0.75\\
\enddata
\tablenotetext{a}{Name formed from J2000 positions (hhmm.mmdddmm.m). }
\tablenotetext{b}{Position of peak surface brightness within clump (accurate to 3\arcsec).}
\tablenotetext{c}{Radius, peak flux, and total flux are derived from {\it clfind} (Williams et al. 1994). {The peak flux and total flux have uncertainties of about 20\,percent, mostly due to absolute flux calibration. The radius has not been deconvolved from the telescope beam.}}
\tablenotetext{d}{Quantities derived from Bonnor-Ebert analysis (see text).}
\tablenotetext{e}{Mass derived from the total flux assuming $T_d = 20\,$K and
$\kappa_{850} = 0.02\,$cm$^{2}$g$^{-1}$.}
\end{deluxetable}





\begin{deluxetable}{lccrr}
\tabletypesize{\footnotesize}
\tablewidth{400pt}
\tablecaption{Submillimeter wavelength properties of clumps in Orion B South \label{t_450850}}
\tablehead{
\colhead{Name\tablenotemark{a}}&
\colhead{$S_{450}/S_{850}$\tablenotemark{b}}&
\colhead{$S_{450}^{\rm peak}/S_{850}^{\rm peak}$\tablenotemark{b}}&
\colhead{$T_d$(BE)\tablenotemark{c}}&
\colhead{$T_d$(SM)\tablenotemark{d}}\\
\colhead{(SMM J)}&
\colhead{}&
\colhead{}&
\colhead{(K)}&
\colhead{(K)}\\
}

\startdata
054086-01528&   24.4&   23.5& 18& 50\\
054089-02271&    8.2&    7.9& 16& 21\\
054090-01531&   20.8&   16.6& 16& 50\\
054091-02278&    9.4&    8.7& 19& 28\\
054096-02074&   10.0&    9.7& 15& 35\\
054097-02086&    8.6&    8.3& 18& 22\\
054100-02092&    8.6&    8.6& 19& 23\\
054105-01558&   12.4&   11.0& 16& 50\\
054112-02273&   10.5&    8.7& 16& 43\\
054116-01580&    2.9&    5.7& 15&  7\\
054140-01507&    8.0&   10.5& 13& 19\\
054141-02180&    7.5&    5.7& 25& 17\\
054141-02289&   13.1&   12.1& 15& 50\\
054142-02160&    8.4&    8.4& 19& 21\\
054142-02194&    5.4&    6.5& 19& 12\\
054146-02208&    6.9&    7.9& 19& 15\\
054146-02239&   11.5&   10.2& 16& 50\\
054148-02200&    6.4&    7.2& 20& 14\\
054149-02212&    8.0&    6.7& 17& 20\\
054149-02232&    8.9&    8.3& 13& 24\\
054151-02251&   12.0&   13.0& 21& 50\\
054152-02262&    7.9&    9.9& 19& 19\\
054153-02189&    2.9&    4.5& 17&  7\\
054155-01497&    8.0&    9.2& 18& 19\\
054156-02209&    6.1&    7.7& 15& 13\\
054156-02250&   13.1&   14.1& 16& 50\\
054158-01558&   11.8&   12.5& 18& 50\\
054158-02303&   11.6&   11.8& 16& 50\\
054159-01514&   13.0&   11.4& 16& 50\\
054160-01492&    6.7&    9.9& 14& 15\\
054160-01564&   15.5&   19.4& 26& 50\\
054160-02300&   13.0&   12.3& 21& 50\\
054161-01495&    9.6&   10.4& 15& 29\\
054162-02172&    6.8&    5.0& 24& 15\\
054163-01509&   13.3&   12.2& 18& 50\\
054163-01558&   17.5&   18.9& 28& 50\\
054164-01504&   10.4&   10.9& 18& 40\\
054167-02171&    8.5&    9.1& 27& 22\\
054170-02162&    6.0&    8.5& 21& 13\\
054172-01543&   14.0&   11.2& 47& 50\\
054173-01527&   15.2&   16.5& 25& 50\\
054173-01547&   13.9&   11.5& 42& 50\\
054173-02170&    6.1&    8.4& 20& 13\\
054174-01556&   12.8&    9.8& 54& 50\\
054175-01560&   16.3&   12.4& 49& 50\\
054182-01576&   22.4&   17.7& 26& 50\\
054182-01596&   19.3&   13.9& 16& 50\\
054185-01534&   18.5&   24.2& 15& 50\\
054188-02002&   17.6&   15.6& 16& 50\\
054190-02004&   17.3&   15.8& 15& 50\\
054192-01565&   23.4&   29.4& 26& 50\\
054195-02009&   15.1&   12.4& 13& 50\\
054204-02077&    7.9&    6.9& 16& 19\\
054205-02025&   11.3&    9.9& 18& 50\\
054207-02017&   13.5&   14.8& 17& 50\\
054207-02032&   10.9&   14.3& 14& 50\\
054218-02042&    7.6&    8.6& 13& 18\\
\enddata
\tablenotetext{a}{Name formed from J2000 positions (hhmm.mmdddmm.m). }
\tablenotetext{b}{The 850\,$\mu$m and 450\,$\mu$m peak flux and total flux are derived from {\it clfind} (Williams et al. 1994) after convolution to an identical beam size.}
\tablenotetext{c}{Quantity derived from Bonnor-Ebert analysis (see text).}
\tablenotetext{d}{Quantity derived from spectral energy fit to 450 and 850$\mu$m integrated fluxes (see text).}
\end{deluxetable}





\begin{deluxetable}{lcccccc}
\tabletypesize{\footnotesize}
\tablewidth{400pt}
\tablecaption{Dust and gas properties of  clumps in NGC\,2023\label{t_2023}}
\tablehead{
\colhead{Name\tablenotemark{a}}&
\colhead{$T_d$(BE)}&
\colhead{$T_d$(SM)}&
\colhead{$T_{\rm gas}$(C$^{18}$O)}&
\colhead{N(850)\tablenotemark{b}}&
\colhead{N(C$^{18}$O)}&
\colhead{N(850)/N(C$^{18}$O)}\\
\colhead{(SMM J)}&
\colhead{(K)}&
\colhead{(K)}&
\colhead{(K)}&
\colhead{($10^{22}\ $cm$^{-2}$)}&
\colhead{($10^{22}\ $cm$^{-2}$)}&
\colhead{}\\
}
\startdata
054141-02180& 25& 19& 18& 12.1& 3.4& 3.6\\
054153-02189& 17&  8& 14&  2.1& 4.2& 0.5\\
054162-02172& 24& 16& 18&  6.6& 4.9& 1.3\\
054167-02171& 27& 24& 25&  3.8&10.0& 0.4\\
054173-02170& 20& 14& 15&  2.6& 5.8& 0.4\\
\enddata
\tablenotetext{a}{Name formed from J2000 positions (hhmm.mmdddmm.m). }
\tablenotetext{b}{Column density derived assuming the BE temperature for the dust.}
\end{deluxetable}

\clearpage

\begin{references}

\reference{Aea02} Abergel, A. et al. 2002, \aap, 389, 239
\reference{Aea03} Abergel, A. et al. 2003, \aap, 410, 577 
\reference{ A82} Anthony-Twarog, B.J. 1982, AJ, 87, 1213
\reference{ Ba89} Barnes, P.J., Crutcher, R.M., Bieging, J.H., Storey,
J.W.V., \& Willner, S.P. 1989, \apj, 342, 883
\reference{ BG98} Barranco, J.A. \& Goodman, A.A.\ 1998, \apj, 504, 207
\reference{ BC04} Basu, S. \& Ciolek, G.E.\ 2004, \apjl, in press
\reference{Bea03} Bik, A., Lenorzer, A., Kaper, L., Comer\'on, F.,
Waters, L.B.F.M., de Koter, A., \& Hanson, M.M. 2003, \aap, 404, 249
\reference{BSW04} Beuther, H., Schilke, P., \& Wyrowski, F. 2004,
\apj, in press
\reference{  B56} Bonnor, W.B. 1956, \mnras, 116, 351
\reference{  C00} Carpenter, J.M. 2000, AJ, 120, 3139
\reference{CBH00} Clarke, C.J., Bonnell, I.A., \& Hillenbrand, L.A. 2000, 
in Protostars and Planets IV, ed. V. Mannings, A. P. Boss \&
S. S. Russell (Tucson: University of Arizona Press), 151
\reference{  E55} Ebert, R. 1955, Zs.Ap., 37, 217
\reference{EEP00} Elmegreen, B.G., Efremov, Y, Pudritz, R.E.,\&
Zinnecker, H.  2000, in Protostars and Planets IV, ed. V. Mannings,
A. P. Boss \& S. S. Russell (Tucson: University of Arizona Press), 179
\reference{  E99} Evans, N.J. II 1999, \araa, 37, 311
\reference{Fea04} Fissel, L, Johnstone, D., Avery, L., \& Mitchell,
G.F. 2005, in preparation
\reference{Fea05} Friesen, R.K., Johnstone, D., Naylor, D.A., \& Davis, G.R.
2005, MNRAS, submitted
\reference{GBL97} Goldsmith, P.F., Bergin, E.A., \& Lis, D.C. 1997,
\apj, 491, 615
\reference{Gea98} Goodman, A.A., Barranco, J.A., Wilner, D.J., \&
Heyer, M.H.\ 1998, \apj, 504, 223
\reference{  G74} Grasdalen G.L. 1974, \apj, 193, 373
\reference{ H98} Hartmann, L. 1998, Accretion Processes in Star
Formation (Cambridge: University Press)
\reference{  H83} Hildebrand, R.H. 1983, QJRAS, 24, 267 
\reference{ HS00} Hogerheijde, M.R. \& Sandell, G. 2000, \apj, 534, 880
\reference{HRG99} Holland, W.S. et al. 1999, \mnras, 303, 659
\reference{JSA02} Jenness, T., Stevens, J.A., Archibald, E.N.,
Economou, F., Jessop, N.E., \& Robson, E.I. 2002, \mnras, 336, 14
\reference{JDK04} Johnstone, D., Di Francesco, J., \& Kirk, H. 2004, \apj
611, L45
\reference{JWG00} Johnstone, D., Wilson, C.D., Moriarty-Schieven, G.,
Gainnakopoulou-Creighton, J., \& Gregersen, E. 2000a (Paper I), \apjs,
131, 505
\reference{JWM00} Johnstone, D., Wilson, C.D., Moriarty-Schieven, G.,
Joncas, G., Smith, G., Gregersen, E., \& Fich, M. 2000b (Paper II),
\apj, 545, 327
\reference{JFM01} Johnstone, D., Fich, M., Mitchell, G.F., \&
Moriarty-Schieven, G. 2001 (Paper III), \apj, 559, 307
\reference{JSD05} J\"orgensen, J.K., Sch\"oier, F.L., \& van Dishoeck,
E.F. 2005, astro-ph/0501623
\reference{KMJ01} Kerton, C.R., Martin, P.G., Johnstone, D., \&
Ballantyne, D.R. 2001, \apj, 552, 601
\reference {Kea96} Kramer, C., Stutzki, J. \& Winnewisser, G. 1996,
\aap, 307, 915
\reference{KSR98} Kramer, C., Stutzki, J., R\"ohrig, R., \&
Corneliussen, U. 1998, \aap, 329, 249
\reference{KTG93} Kroupa, P., Tout, C. \& Gilmore, G. 1993, \mnras,
262,545
\reference{LBS91} Lada, E.A., Bally, J., \& Stark, A.A. 
1991a, \apj, 368, 432
\reference{Lada91b} Lada, E.A., DePoy, D.L., Evans II, N.J., 
\&\ Gatley, I. 1991b, \apj, 371, 171 
\reference{LL03}  Lada, C.J. \& Lada, E.A. 2003, \araa, 41, 57
\reference{Lea96} Launhardt,R., Mezger, P.G., Haslam, C.G.T., Kreysa,
E., Lemke, R., Sievers, A. \& Zylka, R. 1996, \aap, 312, 569
\reference{Lea04} Lenorzer, A., Mokiem, M.R., de Koter, A., \& Puls,
J. 2004, \aap, 422, 275
\reference{ LA01} Lombardi, M. \& Alves, J.\ 2001, \aap, 377, 1023
\reference{ MK04} Mac Low, M. \& Klessen, R.S.\ 2004, Reviews of
Modern Physics, 76, 125
\reference{MWB99} Mangum, J.G., Wootten, A., \& Barsony, M. 1999,
\apj, 526, 845
\reference{ MW97} McKee, C.F. \& Williams, J.P. 1997, \apj, 476, 144
\reference{ MS56} Mestel, L. \& Spitzer, L. 1956, \mnras, 116, 505
\reference{Mea88} Mezger, P.G., Chini, R., Kreysa, E., Wink, J.E., \&
Salter, C.J. 1988, \aap, 191, 44
\reference{ MB94} Miesh \& Bally, J. 1994, ApJ, 429, 645
\reference{M93} Mitchell, G. F. 1993, in Graduate Workshop on Star Formation, 
eds. Arcoragi, J.-P., Bastien, P., and Pudritz, R. 
(Departement de Physique, Universite de Montreal), 81 
\reference{Mea01} Mitchell, G.F., Johnstone, D., Moriarty-Schieven,
G., Fich, M., \& Tothill, N.F.H. 2001, \apj, 556, 215
\reference{Mea00} Mookerjea, B., Ghosh, S.K., Rengarajan, T.N.,
Tandon, S.N. \& Verma, R.P.  2000, \aj, 120, 1954
\reference{Mea04} Mookerjea, B., Kramer, C., Nielbock, M., \& Nyman,
L.-A. 2004, \aap, in press
\reference{MAN98} Motte, F., Andr\'e, P., \& Neri, R. 1998, \aap, 336,
150
\reference{MAW01} Motte, F., Andr\'e, P., Ward-thompson, D., \&
Bontemps, S. 2001, \aap, 372L, 41
\reference{ MC99} Mouschovias, T.C. \& Ciolek, G.E.\ 1999, NATO ASIC
Proc.~540: The Origin of Stars and Planetary Systems, 305
\reference{ N84} Nakano, T.\ 1984, Fundamentals of Cosmic Physics, 9,
139
\reference{Oea02} Onishi, T, Mizuno, A., Kawamura, A., Tachihara, K.,
\& Fukui, Y. 2002, \apj, 575, 950
\reference{Pea03} Pound, M.W., Reipurth, B. \& Bally, J. 2003, \aj,
125, 2108
\reference{Pea01} Price, S.D., Egan, M.P., Carey, S.J., Mizuno, D.R., \&
Kuchar, T.A. 2001, \aj, 121, 2819
\reference{RW05}  Reid, M.A. \& Wilson, C.D. 2005, \apj, in press
\reference{Rei04} Reipurth, B., Rodr\'{i}guez, L.F., Anglada, G., \&
Bally, J. 2004, \aj 127, 1736
\reference{ S55} Salpeter, E.E. 1955, \apj, 121, 161
\reference{Sea99} Sandell, G. et al. 1999, \apj, 519, 236
\reference{Sea01} Sandell, G., Jenness, T., McMullin, J.P. \& Shah,
R.Y. 2001, \baas, 34, 562
\reference{ S85} Scalo, J.M.\ 1985, Protostars and Planets II, 201
\reference{ S86} Scalo, J. 1986, Fundamentals of Cosmic Physics, 11, 1
\reference{ S83} Shu, F.H.\ 1983, \apj, 273, 202
\reference{SAL87} Shu, F.H., Adams, F.C., \& Lizano, S.\ 1987, \araa,
25, 23
\reference{TDE00} van der Tak, F.F.S., van Dishoeck, E.F., Evans,
N.J. II, \& Bakker, E.J.. 1999, \apj, 522, 991
\reference{Tea04} Tafalla, M., Myers, P.C., Caselli, P., \& Walmsley,
C.M. 2004, \aap, 416, 191
\reference{ TS98} Testi, L. \& Sargent, A.I. 1998, \apj, 508, L91
\reference{Tea04} Teyssier, D., Foss\'{e}., D., Gerin, M., Pety, J.,
Abergel, A. \& Roueff, E.  2004, \aap, 417, 135
\reference{Vea98} Visser, A.E., Richer, J.S., Chandler, C.J., \&
Padman, R. 1998, \mnras, 301, 585
\reference{WBM00} Williams, J.P., Blitz, L, \& McKee, C.F. 2000, in
Protostars and Planets IV, ed. V. Mannings, A. P. Boss \&
S. S. Russell (Tucson: University of Arizona Press), 97
\reference{WGB94} Williams, J.P., de Geus, E.,J., \& Blitz L. 1994,
\apj, 428, 693
\reference{ WM97} Williams, J.P. \& McKee, C.F. 1997, \apj, 476, 166
\end{references}
\end{document}